\documentclass[11pt]{article}
\usepackage{amssymb}
\usepackage{amsmath}

\newcommand{\beq}{\begin{equation}}
\newcommand{\eeq}{\end{equation}}
\newcommand{\bdm}{\begin{displaymath}}
\newcommand{\edm}{\end{displaymath}}
\newcommand{\beqr}{\begin{eqnarray}}
\newcommand{\eeqr}{\end{eqnarray}}
\newcommand{\beqrn}{\begin{eqnarray*}}
\newcommand{\eeqrn}{\end{eqnarray*}}
\def\l{\lambda}
\def\bchi{\boldsymbol{\chi}}

\begin{document}
\title{Some results on the eigenfunctions of the quantum trigonometric 
Calogero-Sutherland model related to the Lie algebra $E_6$}
\author{J. Fern\'andez N\'u\~{n}ez$^{\;\dagger}$, W. Garc\'{\i}a Fuertes$^{\;
\dagger}$,  A.M. Perelomov$^{\;\ddagger\;}$
\footnote{On leave of absence from the Institute for Theoretical 
and Experimental Physics, 117259, Moscow, Russia. 
Current E-mail address: perelomo@dftuz.unizar.es}\\
{\normalsize {\it $^\dagger$ Dep. de F\'{\i}sica, 
Facultad de Ciencias, Universidad de Oviedo, E-33007 Oviedo, Spain}} \\
{\normalsize {\it $^\ddagger$ Dep. de F\'{\i}sica Te\'orica, 
Facultad de Ciencias, Universidad de Zaragoza, E-50009 Zaragoza, Spain}}}
\date{}
\maketitle
\begin{abstract}\noindent
The quantum trigonometric Calogero-Sutherland 
models related to Lie algebras admit a parametrization in which the dynamical 
variables are the characters of the fundamental representations
of the algebra. We develop here this approach for the case of the exceptional Lie algebra $E_6$.
\end{abstract}
\section{Introduction}
The so-called Calogero-Sutherland or Calogero-Moser models were introduced by Calogero \cite{ca71}, who studied, from the quantum standpoint, the dynamics on the infinite line of a set of pairwise interacting particles through rational plus quadratic potentials, and found that the problem was exactly solvable. Soon afterwards, Sutherland \cite{su72} arrived to similar results for the quantum problem on the circle, this time with trigonometric interaction, and Moser \cite{mo75} showed that the classical version of both models enjoyed integrability in the Liouville sense. The identification of the general scope of these discoveries came with the works of Olshanetsky and Perelomov \cite{op76}-\cite{op78}, who realized that it was possible to associate models of this kind to all the root systems of the simple Lie algebras, and that all these models were integrable, both in the classical and in the quantum framework \cite{op81,op83}. Nowadays, there is a widespread interest in this type of integrable systems, and many mathematical and physical applications for them have been found, see for instance \cite{dv00}.

The Calogero-Sutherland Hamiltonian associated to the root system of a simple Lie algebra $L$ can be written as a second-order differential operator whose variables are the characters of the fundamental representations of the algebra.  As it was shown in the papers \cite{pe98a,prz98,pe99}, and later in \cite{pe00,flp01,fp02,ffp03,ffp03b, ffp05}, this approach gives the possibility of developping some systematic procedures to solve the Schr\"{o}dinger equation and determine important properties of the eigenfunctions, such as recurrence relations or generating functions for some subsets of them. For the moment, the approach has been used only for classical algebras of $A_n$ and $D_n$ type, and recently \cite {ffp05} for the exceptional algebra $E_6$ for a  special value of the coupling constant for which the eigenfunctions are proportional to the characters of the ireducible representations of the algebra. The aim of this paper is to show how to generalize the treatment of \cite{ffp05} to arbitrary values of the coupling constant and to extend some of the particular results found there to the general case.
\section{The Calogero-Sutherland model for $E_6$ in $z$-variables}
The Hamiltonian operator for the trigonometric Calogero-Sutherland model related to the root system of a simple Lie algebra has the generic form
\[
H=\frac{1}{2}(p,p)+\sum_{\alpha\in{\cal R}^+}\kappa_\alpha(\kappa_\alpha-1)\sin^{-2}(\alpha,q),
\]
where $\cal R^+$ is the set of positive roots, $q$ and $p$ are vectors of dimension  $r= \,$rank of the algebra, $(\ ,\ )$ is the usual euclidean scalar product in ${{\bf R}}^r$, and the coupling constants $\kappa_\alpha$ are such that $\kappa_\alpha=\kappa_\beta$ if $||\alpha||=||\beta||$. In particular, because $E_6$ is simply-laced (for all details about the structure of $E_6$ needed to follow the main text, see Appendix A), the Calogero-Sutherland model associated to $E_6$ depends only on one coupling constant $\kappa$. To write $H$ in a more explicit way, it is convenient to use the orthonormal basis $\{e_i,\ i=1,\dots,6\}$ which is related to the generating system of the Appendix A through $e_i=\varepsilon_i-\frac{1}{6}\sum_{j=1}^6\varepsilon_j$.  The expression of $q$ and $p$ in this basis is simply  $q=\sum_{i=1}^6 q_i\, e_i$,  $p=\sum_{i=1}^6 p_i\, e_i$, while
the simple roots are given by:
\beqrn
\alpha_1&=&e_1-e_2\\
\alpha_2&=&\frac{1}{2}\left(-1+\frac{\sqrt{3}}{3}\right)\sum_{j=1}^3 e_j+\frac{1}{2}\left(1+\frac{\sqrt{3}}{3}\right)\sum_{j=4}^6 e_j\\
\alpha_k&=&e_{k-1}-e_k,\ \ \ \ \ k=3,4,5,6.
\eeqrn
The $q$ coordinates are assumed to take values in the interval $[0,\pi]$, and therefore the Hamiltonian can be interpreted as describing the dynamics of a system of six particles moving on the circle, but notice that there is not translational invariance. We recapitulate some important facts about this model which follow from the general structure of the quantum Calogero-Sutherland models related to Lie algebras \cite{op83}. The ground state energy and (non-normalized) wave function are
\begin{eqnarray*}
E_0(\kappa)&=&2 (\rho, \rho)\kappa^2= 156\kappa^2\nonumber\\
\Psi_0^\kappa(q)&=&{\prod_{\alpha\in {\cal R}^+}\sin^\kappa(\alpha, q)},
\end{eqnarray*}
with $\rho$ being the Weyl vector, while the excited states depend on a six-tuple of quantum numbers ${\bf m}=(m_1,m_2,m_3,m_4,m_5,m_6)$, and satisfy the Schr\"odinger equation
\begin{eqnarray}
H\Psi^\kappa_{\bf m}&=&E_{\bf m}(\kappa)\Psi_{\bf m}^\kappa\nonumber\\
E_{\bf m}(\kappa)&=&2 (\lambda+\kappa\rho,\lambda+\kappa\rho),\label{105}
\end{eqnarray}
where $\lambda$ is the highest weight of the irreducible representation of $E_6$ labelled by ${\bf m}$, i. e. $\lambda=\sum_{i=1}^6 m_i \lambda_i$. By substitution in (\ref{105}) of
\beq
\Psi_{\bf m}^\kappa(q)=\Psi_0^\kappa(q)\Phi_{\bf m}^\kappa(q),\label{107}
\eeq
we are led to the eigenvalue problem
\beq
-\Delta^\kappa\Phi_{\bf m}^\kappa=\varepsilon_{\bf m}(\kappa)\Phi_{\bf m}^\kappa
\label{sch}
\eeq
with
\beq
\Delta^\kappa=\frac{1}{2}\Delta+\kappa\sum_{\alpha\in {\cal R}^+}  {\rm ctg}(\alpha, q)(\alpha,\nabla_q)
\label{4b},
\eeq
and
\beq
\label{energ}
\varepsilon_{\bf m}(\kappa)=E_{\bf m}(\kappa)-E_0(\kappa)= 2(\lambda, \lambda+2\kappa\rho).
\eeq
Taking into account that $A_{jk}^{-1}= (\lambda_j,\lambda_k)$, it is possible to give a more explicit expression for $\varepsilon_{\bf m}(\kappa)$:
\beq
\varepsilon_{\bf m}(\kappa)=2\sum_{j,k=1}^6 A_{jk}^{-1} m_j m_k+4\kappa\sum_{j,k=1}^6 A_{jk}^{-1}m_j.\label{eigenvalues}
\eeq

Now the main problem is to solve (\ref{sch}). As it has been shown for other algebras \cite{pe98a,prz98,pe99,ffp03}, the best way to do this  is to use a set of independent variables which are invariant under the Weyl symmetry of the Hamiltonian, namely the characters $z_k, k=1,\ldots,6$, of the six fundamental representations of the Lie algebra $E_6$. We can infer from (\ref{4b}) the structure of $\Delta^\kappa$ when written in the $z$-variables:
\beq
\Delta^\kappa=\sum_{j, k=1}^6 a_{jk}(z)\partial_{z_j}\partial_{z_k}+\sum_{j=1}^6 \left[b_j^{(0)}(z)+\kappa b_j^{(1)}(z)\right]\partial_{z_j}
\label{structure}.
\eeq

As a matter of fact, the eigenfunctions of $\Delta^{(0)}$ and $\Delta^{(1)}$ are (proportional to) the monomial symmetric functions $M_\lambda=\sum_{s\in W}\exp[i(s\lambda,q)]$ ($W$ is the Weyl group) and the characters $\bchi_\lambda$ of the irreducible representations of the algebra $E_6$, respectively \cite{op83}. Thus,  knowing the characters $z_i=\bchi_{\lambda_i}$ of the fundamental representations and the products $z_i\,z_j$ through the Clebsch-Gordan series for the  algebra, we are able to find the Hamiltonian $\Delta^{(1)}$, that is, we obtain  the coefficients $a_{jk}(z)$ entering in the expression of all the Hamiltonians $\Delta^\kappa$ and also the coefficients $b_j^{(0)}(z)+ b_j^{(1)}(z)$. In the previous paper \cite{ffp05} we computed the needed Clebsch-Gordan series and showed these coefficients.

On the other hand, knowing  enough monomial symmetric functions in terms of the fundamental characters, $M_\lambda(z)$,  we can complete the form of $\Delta^\kappa$, for we know that
\beq
\label{bes}
\Delta^{(0)}M_\lambda=\varepsilon_{\bf m}(0)M_\lambda=2(\lambda,\lambda)M_\lambda,
\eeq
a system of linear equations which can be solved  for the  coefficients $b_j^{(0)}(z)$, $j=1,\dots, 6$. To this end,  remind 
that the characters can be expanded as sums of monomial functions (\cite{aram}),
\bdm
\bchi_\lambda=M_\lambda+a_1\, M_{\mu_1}+a_2\, M_{\mu_2}+\cdots,
\edm
where the set of $\mu_k$ entering in the expansion is easy to determine: they are the dominant weights such that $\mu_i=\lambda-\sum_{j=1}^6 n_j \alpha_j$ with $n_j\geq 0$ and $(\mu_i,\alpha_k)\geq 0$ for the six simple roots $\alpha_k$. The coefficients $a_k$, on the other hand, represent the multiplicities of the weights $\mu_i$ in the representation with highest weight $\lambda$. Here it will suffice to deal with the following expansions:
\beqr
\bchi_{100000}^{(27)}&=& z_1 = M_{100000}^{(27)}\,,\nonumber\\
\bchi_{010000}^{(78)}&=& z_2 = M_{010000}^{(72)}+a M_{000000}^{(1)}\,,\nonumber\\
\bchi_{001000}^{(351)}&=& z_3 = M_{001000}^{(216)}+b M_{000001}^{(27)}\,,\nonumber\\
\bchi_{000100}^{(2925)}&=& z_4 = M_{000100}^{(720)}+c M_{100001}^{(270)}+d M_{010000}^{(72)}+e M_{000000}^{(1)}\,,\label{expansion}\\
\bchi_{000010}^{(351)}&=& z_5 = M_{000010}^{(216)}+b M_{100000}^{(27)}\,,\nonumber\\
\bchi_{000001}^{(27)}&=& z_6 = M_{000001}^{(27)}\,,\nonumber\\
\bchi_{100001}^{(650)}&=& z_1 z_6-z_2-1=M_{100001}^{(270)}+f M_{010000}^{(72)}+ g M_{000000}^{(1)}\,,\nonumber
\eeqr
where the form of $\bchi_{100001}$ comes from the list in \cite{ffp05} and the numbers appearing in parentheses as superscripts are either the dimensions of the representations or the dimensions of the linear spaces generated by the orbits of the Weyl group corresponding to the monomial functions. The former can be computed from the Weyl dimension formula, while the latter follow easily from the fact that the Weyl group of the subalgebra of $E_6$ obtained by removing from the Dynkin diagram the dots corresponding to the weight defining the monomial function acts trivially on such a weight: for instance, removing the dot associated to $\lambda_1$ we obtain the Dynkin diagram of $D_5$, and hence 
\bdm
\dim M_{100000}=\frac{|W_{E_6}|}{|W_{D_5}|}=\frac{2^7\cdot 3^4\cdot 5}{\frac{1}{2} 10!!}=27,
\edm
and so on. 

While the dimensions shown in (\ref{expansion}) suffice for fixing $a=6$, $b=5$, for the computation of remaining multiplicities we need to use the Freudenthal formula \cite{ov90}
\beq
n_\mu=\frac{\sum_{\alpha >0} \sum_{k=1}^\infty 2 n_{\mu+k \alpha} (\mu+k\alpha,\alpha)}{(\lambda+\mu+2\rho,\lambda-\mu)}.\label{freud}
\eeq
Here $n_\mu$ stands for the multiplicity of the weight $\mu$ in the representation of highest weight $\lambda$, the first sum extends over positive roots, and $\rho$ is the Weyl vector. The application of the Freudenthal formula is quite easy for the representations at stake. Let us see, for instance, how to compute $c$ in (\ref{expansion}). In this case $\lambda=\lambda_4$ and $c=n_\mu=n_{\lambda_1+\lambda_6}$. The scalar product of the vector $\beta=\lambda-\mu=\alpha_2+\alpha_3+2\alpha_4+\alpha_5$ with $\mu$ is $(\beta,\mu)=0$ and, due to the fact that the length of $\mu+\alpha$, with $\alpha$ a positive root, is $|\mu+\alpha|=(|\mu|^2+2+2(\mu,\alpha))^{\frac{1}{2}}$, the only roots entering in (\ref{freud}) are the positive roots $\alpha$ such that $(\mu,\alpha)=0$, because otherwise $|\mu+\alpha|>|\mu+\beta|=|\lambda|$ and $\mu+\alpha$ would lie outside of the weight diagram for the representation ${R}_\lambda$. Looking at the table of positive roots in Appendix A, we check that there are 12 of them with $(\mu,\alpha)=0$. For all of these, $\mu+\alpha$ lies on the orbit of $\lambda$, and thus $n_{\mu+\alpha}=1$. This gives
\bdm
c=n_{\lambda_1+\lambda_6}=12\frac{2(\mu,\alpha)+2(\alpha,\alpha)}{(\lambda_4+\lambda_1+\lambda_6+2\sum_{i=1}^6 \lambda_i,\alpha_1+\alpha_3+2\alpha_4+\alpha_5)}=12\frac{2\cdot 0+2\cdot 2}{2+2\cdot 5}=4.
\edm
To compute $d$ we proceed much in the same way. Now $\mu=\lambda_2$ and $\beta=\lambda-\mu=\alpha_1+\alpha_2+2\alpha_3+3\alpha_4+2\alpha_5+\alpha_6$. It follows that $(\mu,\beta)$=1, and thus only positive roots with $(\mu,\alpha)=0$ or $(\mu,\alpha)=1$ enter in the Freudenthal formula. There are 20 positive roots with $(\mu,\alpha)=1$, and for them $n_{\mu+\alpha}$=1 because they are in the orbit of $\lambda$. The number of positive roots with $(\mu,\alpha)=0$ is 15, and for them $\mu+\alpha$ lies in the orbit of $\lambda_1+\lambda_6$, so that their multiplicities are $n_{\mu+\alpha}=4$. This gives $d=n_{\lambda_2}=15$. Once we know $c$ and $d$, we compute $e$ by balancing dimensions in (\ref{expansion}), and obtain $e=45$. A similar use of the Freudenthal formula gives $f=5$, and therefore $g=20$.

With all the coefficients in (\ref{expansion}) being fixed, we can now solve for the monomial functions corresponding to the fundamental weights. We find
\beqrn
M_{100000}&=&z_1\,,\\
M_{010000}&=&z_2-6\,,\\
M_{001000}&=&z_3-5 z_6\,,\\
M_{000100}&=&z_4-4 z_1 z_6+9z_2+9\,,\\
M_{000010}&=&z_5-5 z_1\,,\\
M_{000001}&=&z_6.
\eeqrn
The remaining step is to substitute these monomials in (\ref{bes}) and to solve  the linear system  for the coefficients $b_j^{(0)}(z)$; 
the outcome is
\bdm
\begin{array}{llll}
b_1^{(0)}=\frac{8}{3}z_1,\ &b_2^{(0)}=4z_2-6,\ &b_3^{(0)}=\frac{20}{3}z_3-20z_6,\ & \\
b_4^{(0)}=12z_4-16z_1z_6-24z_2+36,\ &b_5^{(0)}=\frac{20}{3}z_5-20z_1,\ & b_6^{(0)}=z_6.
\end{array}
\edm

With this and the form of $\Delta^{(1)}$ given in \cite{ffp05}, we can now write the full set of   coefficients in (\ref{structure}):
\beqrn
a_ {11}(z)&=&\frac{8}{3}z_1^2 - 4z_3 - 20z_6,  \\ \nonumber
a_ {12} (z)&=&2 z_1 z_2-26 z_1  - 10 z_5,  \\ \nonumber
a_ {13} (z)&=& \frac{10}{3}z_1z_3+18 - 12z_2  - 6z_4 - 18z_1z_6,  \\ \nonumber
a_ {14} (z)&=&  4z_1z_4+ 18z_1 - 10z_1z_2  - 18z_5 - 8z_2z_5 - 8z_3z_6 + 
    8z_6^2,  \\ \nonumber
a_ {15} (z)&=&   \frac{8}{3}z_1z_5 -10 z_3   - 26z_6 - 10z_2z_6,  \\ \nonumber
a_ {16} (z)&=&    \frac{4}{3}z_1z_6 -36  - 12z_2,    \\ \nonumber
a_ {22} (z)&=&    2z_2^2-18  - 6z_2  - 2z_4 - 8z_1z_6,   \\ \nonumber
a_ {23} (z)&=&    4z_2z_3 -24 z_1^2 + 14z_3  - 8z_1z_5 - 2z_6 - 10z_2z_6,  \\ \nonumber
a_ {24} (z)&=&    6z_2z_4-18 z_2 - 12z_2^2 - 10z_1z_3 + 24z_4  - 6z_3z_5 + 26z_1z_6 - 8 z_1z_2z_6 - 10z_5z_6, \\ \nonumber
  a_ {25} (z)&=&   4z_2z_5  -2 z_1 - 10z_1z_2 + 14z_5  - 8z_3z_6 - 24z_6^2 , \\ \nonumber
  a_ {26} (z)&=& 2 z_2 z_6-10 z_3 - 26 z_6,  \\ \nonumber
  a_ {33} (z)&=&   \frac{10}{3}z_3^2+ 14z_1 - 12z_1z_2   - 2z_1 
    z_4 + 16z_5 - 4z_2z_5 - 8z_1^2z_6 + 4z_3z_6 - 6z_6^2 , \\ \nonumber
  a_ {34} (z)&=&   8z_3 z_4+10z_1^2 - 10z_1^2z_2 + 18z_3 - 2z_2z_3  - 6z_1z_2z_5 - 10z_5^2 - 18z_6 + 8z_2z_6 - 10z_2^2z_6 \\ \nonumber
  &&   -  8z_1z_3z_6 +20z_4z_6 + 8z_1z_6^2,  \\ \nonumber
  a_ {35} (z)&=&    \frac{16}{3}z_3 z_5  -36  + 24z_2 - 12z_2^2 - 10z_1z_3 + 24z_4  - 16z_1z_6 - 8z_1z_2z_6 - 10z_5z_6,  \\ \nonumber
  a_ {36} (z)&=&     \frac{8}{3}z_3z_6-26 z_1 - 10z_1z_2 - 10z_5 ,   \\ \nonumber
  a_ {44} (z)&=&    6 z_4^2 -4 z_1^3 - 6z_2^3 + 18z_1z_3 - 6z_1z_2z_3 - 18z_4 + 18z_2z_4  + 8z_1^2z_5 - 18z_3z_5 - 2z_2z_3z_5 -  4z_1z_5^2\\ \nonumber
   && -18z_1z_6 + 14z_1z_2z_6 - 4z_1z_2^2z_6 - 4z_3^2z_6 + 
    8z_1z_4z_6 + 18z_5z_6 - 6z_2z_5z_6 + 8z_3z_6^2 - 4z_6^3,  \\ \nonumber
    a_ {45} (z)&=&   8z_4z_5  -18 z_1 + 8 z_1z_2 - 10z_1z_2^2 - 10z_3^2 + 20z_1z_4 + 18z_5 - 2z_2z_5+ 8z_1^2z_6 - 6z_2z_3z_6 \\Ê\nonumber
&&  -  8z_1 z_5z_6 +10z_6^2 - 10z_2z_6^2,  \\ \nonumber
  a_ {46} (z)&=&  4 z_4z_6 +8z_1^2 - 18z_3 - 8z_2z_3 - 8z_1z_5 + 18z_6 - 10z_2z_6,    \\ \nonumber
  a_ {55} (z)&=& \frac{10}{3}z_5^2   -6 z_1^2 + 16z_3 - 4z_2z_3 + 4z_1z_5 +   14z_6 - 12z_2z_6 - 2z_4z_6 - 8z_1z_6^2,  \\ \nonumber
  a_ {56} (z)&=&   \frac{10}{3}z_5z_6 +18 - 12z_2 - 6z_4 - 18z_1z_6,    \\ \nonumber
  a_ {66} (z)&=&   \frac{4}{3}z_6^2  -10 z_1 - 2z_5,   \\
b_1(z)&=&b_1^{(0)}(z)+\kappa b_1^{(1)}(z)=\Big(32\kappa+\frac{8}{3}\Big)z_1,\nonumber\\
b_2(z)&=&b_2^{(0)}(z)+\kappa b_2^{(1)}(z)=(44\kappa+4)z_2+24(\kappa-1),\nonumber\\
b_3(z)&=&b_3^{(0)}(z)+\kappa b_3^{(1)}(z)=\Big(60\kappa+\frac{20}{3}\Big)z_3+20(\kappa-1) z_6,\nonumber\\
b_4(z)&=&b_4^{(0)}(z)+\kappa b_4^{(1)}(z)=(84\kappa+12)z_4+(\kappa-1)(16z_1 z_6+24z_2-36),\nonumber\\
b_5(z)&=&b_5^{(0)}(z)+\kappa b_5^{(1)}(z)=\Big(60\kappa+\frac{20}{3}\Big)z_5+20(\kappa-1) z_1,\nonumber\\
b_6(z)&=&b_6^{(0)}(z)+\kappa b_6^{(1)}(z)=\Big(32\kappa+\frac{8}{3}\Big)z_6.\nonumber
\eeqrn
\section{Computation of polynomials and deformed Clebsch-Gordan series}
The eigenfunctions $\Phi_{\bf m}^\kappa(q)$ are polynomials when expressed in  $z$ variables, $\Phi_{\bf m}^\kappa(q)=P_{\bf m}^\kappa(z)$. The Schr\"{o}\-din\-ger equation can then be solved by applying a systematic procedure, which is suitable to be implemented in a computer program able to carry out symbolic calculations. We propose two alternative methods to find the Schr\"{o}dinger eigenfunctions:
\begin{enumerate}
\item Given a weight $n_1\lambda_1+n_2\lambda_2+n_3\lambda_3+n_4\lambda_4+n_5\lambda_5+n_6\lambda_6$, let us denote $z^{\bf n}=z_1^{n_1}z_2^{n_2}z_3^{n_3}z_4^{n_4}z_5^{n_5}z_6^{n_6}$. Thus, $\Delta^{\kappa}$ acting on $z^{\bf n}$ gives 
\beq
\Delta^{\kappa} z^{\bf n}=\sum_{\beta\in\Lambda} k_{\beta,{\bf n}} (\kappa) \, z^{\bf n-\beta},
\label{e64}
\eeq
where $\Lambda$ includes  only integral linear combinations of the simple roots with non-negative coefficients   and, of course, in the exponent of (\ref{e64}) we express $\beta$ in the basis of fundamental weights. In particular, $k_{0,{\bf n}}(\kappa)=\varepsilon_{\bf n}(\kappa)$. The polynomials $P_{\bf m}(z)$ can be written as
\[
P_{\bf m}^\kappa(z)=\sum_{\mu \in Q^+({\bf m})} c_\mu(\kappa) z^{\bf m-\mu}, \ \ \ \ c_0=1,
\]
where again the $\mu$ in $Q^+({\bf m})$ are integral linear combinations of the simple roots with non-negative coefficients such  that  they do not give rise to negative powers of the $z$'s. By substituting in the Schr\"{o}dinger equation we find the iterative formula
\[
c_\mu(\kappa)=\frac{1}{\varepsilon_{\bf m}(\kappa)-\varepsilon_{\bf m-(\mu-\beta)}(\kappa)}\sum_{\beta\in\Lambda,\beta\neq 0}k_{\beta-{\bf m}-(\mu-\beta)}(\kappa)\, c_{\mu-\beta}(\kappa).
\]
To use this formula in practice, one should take into account the heights of the $\mu's$, because each coefficient $c_\mu$ can depend only on some of the $c_\nu$ such that ${\rm ht}(\nu)<{\rm ht}(\mu)$.

\item The product $z_1^{m_1}z_2^{m_2}z_3^{m_3}z_4^{m_4}z_5^{m_5}z_6^{m_6}$ can be expanded on the basis of the orthogonal Polynomials $P_{\bf m}^\kappa(z)$ as
\[
z_1^{m_1}z_2^{m_2}z_3^{m_3}z_4^{m_4}z_5^{m_5}z_6^{m_6}=P_{\bf m}^\kappa(z) +\sum_{\beta\in S_{\bf m}}n_{\beta}(\kappa) P_{\bf m-\beta}^\kappa(z)\,.
\]
In each particular case, it is not difficult to elaborate a list with all the elements in $S_{\bf m}$ (they are the same integral dominant weights which appear in the corresponding Clebsch-Gordan series, see \cite{ffp05}). Furthermore, the operator $\Delta^{\kappa}-\varepsilon_{\bf n}(\kappa)$ annihilates the character $P_{\bf n}^\kappa$. Taking this into account, we can obtain the eigenfunctions using the formula
\[
P_{\bf m}^\kappa=\Big\{\prod_{\beta \in S_{\bf m}}\left(\Delta^{\kappa}-\varepsilon_{\bf m-\beta}(\kappa)\right)\Big\} z^{\bf m}.
\]
\end{enumerate}
Through any of these methods, it is possible to compute the characters rather quickly. As an illustration, we offer a list of polynomials and monomial functions in Appendix B. For a similar list of characters, see \cite{ffp05}.

Once we have a method for the computation of the polynomials, we can extend it to produce an algorithm for calculating deformed Clebsch-Gordan series for the product of them. Suppose that we want to obtain the series for $P_{\bf m}^\kappa\cdot P_{\bf n}^\kappa$. We  list the possible dominant weights entering in the series arranged by heights
\[
P_{\bf m}^\kappa \cdot P_{\bf n}^\kappa=P_{\bf m+n}^\kappa+n_{\mu_1}(\kappa) P_{\mu_1}^\kappa+n_{\mu_2}(\kappa) P_{\mu_2}^\kappa+\ldots
\]
The coefficient  $n_{\mu_1}(\kappa)$ is simply the difference between the coefficients of $z^{\bf \mu_1}$ in $P_{\bf m}^\kappa\cdot P_{\bf n}^\kappa$ and in $P_{\bf m+n}^\kappa$. Then, $n_{\mu_2}(\kappa)$ is the difference between the coefficient of $z^{\bf \mu_2}$ in $P_{\bf m}^\kappa\cdot P_{\bf n}^\kappa$ and the sum of the corresponding coefficients in $P_{\bf m+n}^\kappa$ and $P_{\bf \mu_1}^\kappa$, and so on. As an example, we present a list with all the cuadratic deformed Clebsch-Gordan series in Appendix C.
\section{Some recurrence relations}
The approach we are describing is also useful to find the form of the recurrence relations for products $z_j P_{\bf m}^\kappa(z)$. Considered in full generality, these recurrence relations are extremely complicated, but for some special cases they can be written in explicit form. Let us consider, for instance, the recurrence relation for $z_1 P_{n\lambda_1}^\kappa$ with arbitrary $n$. If we express the weights of the representation $R_{\lambda_1}$ (which are all the combinations $\varepsilon_i\pm\varepsilon$, $-\varepsilon_i-\varepsilon_j$) in the basis of fundamental weights, we see that there are only three whose coefficients for $\lambda_i$, $i\neq 1$, are all non-negative, namely $\lambda_1, -\lambda_1+\lambda_3$ and $-\lambda_1+\lambda_6$. Hence, the form of the series should be
\beq
z_1 P^\kappa_{n00000}=P^\kappa_{(n+1)00000}+ a_n(\kappa)P^\kappa_{(n-1)01000}+b_n(\kappa) P^\kappa_{(n-1)00001}, \label{ser11}
\eeq
where we have to fix $a_n(\kappa)$ and $b_n(\kappa)$. Now,  solving the Schr\"{o}dinger equation by means of the first of the two methods described in Sect. 3, one finds
\begin{eqnarray*}
P^\kappa_{n00000}&=&z_1^{n}+\frac{(1-n) n}{n+\kappa-1}z_1^{n-2} z_3+\frac{(1-n)n\kappa(n+5\kappa-2)}{(n+\kappa-1)(n+\kappa-2)(n+4\kappa-1)}z_1^{n-2}z_6+\ldots,\\
P^\kappa_{(n-1)01000}&=&z_1^{n-1} z_3+\frac{10 \kappa^3-5(1+3n)\kappa^2-2(-1-9n+5n^2)\kappa-n(2-3n+n^2)}{(n+\kappa-2)(n+\kappa-1)(n+7\kappa)}z_1^{n-1} z_6+\ldots,\\
P^\kappa_{(n-1)00001}&=&z_1^{n-1} z_6+\ldots.
\end{eqnarray*}
Substituting in (\ref{ser11}), we can solve for $a_n(\kappa)$ and $b_n(\kappa)$ with the results
\beqrn
a_n(\kappa)&=&\frac{n(n+2\kappa-1)}{(n+\kappa)(n+\kappa-1)},\\
b_n(\kappa)&=&\frac{n(n+3\kappa)(n+5\kappa-1)(n+8\kappa-1)}{(n+\kappa-1)(n+4\kappa-1)(n+4\kappa)(n+7\kappa)}.
\eeqrn

We list below the series of the form $z_1 P^\kappa_{n\lambda_k}$ obtained through the same procedure: 
\begin{eqnarray*}
&&z_1 P^\kappa_{0n0000}=P^\kappa_{1n0000}+ c_n(\kappa) P^\kappa_{0(n-1)0010}+d_n(\kappa) P^\kappa_{1(n-1)0000},\\
&&z_1 P^\kappa_{00n000}=P^\kappa_{10n000}+ e_n(\kappa)  P^\kappa_{00(n-1)100}+ f_n(\kappa) P^\kappa_{10(n-1)001}+g_n(\kappa) P^\kappa_{10(n-1)001},\\
&&z_1 P^\kappa_{000n00}=P^\kappa_{100n00}+h_n(\kappa)  P^\kappa_{010(n-1)10}+i_n(\kappa)  P^\kappa_{001(n-1)01}+j_n(\kappa) P^\kappa_{110(n-1)00}+ k_n(\kappa)  P^\kappa_{000(n-1)10},\\
&&z_1 P^\kappa_{0000n0}=P^\kappa_{1000n0}+ l_n(\kappa)  P^\kappa_{0100(n-1)1}+p_n(\kappa) P^\kappa_{0010(n-1)0}+q_n(\kappa) P^\kappa_{0000(n-1)1},\\
&&z_1 P^\kappa_{00000n}=P^\kappa_{10000n}+ r_n(\kappa) P^\kappa_{01000(n-1)}+s_n(\kappa) P^\kappa_{00000(n-1)}\,,
\end{eqnarray*}
where
\beqrn
c_n(\kappa)&=&\frac { n (-1 + n + 5 \kappa) }{ (-1 + n + \kappa) (n + 4 \kappa) }, \\
d_n(\kappa)&=&\frac { 2 n (n + 2 \kappa) (-1 + n + 6 \kappa) (-1 + n + 8 \kappa) (-1 + 2 n + 12 \kappa) }{ 
  (-1 + n + \kappa) (-1 + n + 3 \kappa) (n + 7 \kappa) (-1 + 2 n + 11 \kappa) (2 n + 11 \kappa) }, \\
e_n(\kappa)&=&\frac { n (-1 + n + 3 \kappa) }{ (-1 + n + \kappa) (n + 2 \kappa) } ,\\
f_n(\kappa)&=&\frac { 2 n (n + 2 \kappa) (-1 + n + 4 \kappa) (-1 + n + 6 \kappa) (-1 + 2 n + 8 \kappa) }{ 
  (-1 + n + \kappa) (-1 + n + 3 \kappa) (n + 5 \kappa) (-1 + 2 n + 7 \kappa) (2 n + 7 \kappa) }, \\
 g_n(\kappa)&=&\frac { n (n + \kappa) (n + 3 \kappa) (-1 + n + 5 \kappa) (-1 + n + 6 \kappa) (-1 + 2 n + 11 \kappa) }{ 
  (-1 + n + \kappa) (-1 + n + 2 \kappa) (n + 4 \kappa) (n + 5 \kappa)^2 (-1 + 2 n + 7 \kappa) }, \\
 \eeqrn
  \beqrn
h_n(\kappa)&=&\frac { n (-1 + n + 4 \kappa) }{ (-1 + n + \kappa) (n + 3 \kappa) }, \\
i_n(\kappa)&=&\frac { n (n + \kappa) (-1 + n + 3 \kappa) (-1 + n + 4 \kappa) (-1 + 2 n + 7 \kappa) }{ 
  (-1 + n + \kappa) (-1 + n + 2 \kappa) (n + 3 \kappa)^2 (-1 + 2 n + 5 \kappa) }, \\
j_n(\kappa)&=&\frac { 6 n (n + \kappa) (n + 2 \kappa) (-1 + n + 4 \kappa) (-1 + n + 5 \kappa) (-1 + 2 n + 8 \kappa) 
   (-1 + 3 n + 11 \kappa) }{ (-1 + n + \kappa) (-1 + n + 2 \kappa) (n + 4 \kappa) (-1 + 2 n + 5 \kappa) 
   (2 n + 7 \kappa) (-1 + 3 n + 10 \kappa) (3 n + 10 \kappa) }, \\
k_n(\kappa)&=&\frac { 3 n (n + \kappa) (n + 2 \kappa) (-1 + n + 4 \kappa) (-1 + n + 5 \kappa) (2 n + 5 \kappa) 
   (-1 + 2 n + 8 \kappa)}{ 
  (-1 + n + \kappa) (-1 + n + 2 \kappa) (n + 4 \kappa)^2 (-1 + 2 n + 5 \kappa) (-1 + 2 n + 6 \kappa) 
   (2 n + 7 \kappa)} \\ &\times&\frac{ (-1 + 2 n + 9 \kappa)(-1+3n+12\kappa)}{(-1 + 3 n + 10 \kappa)(3n+11)\kappa},\\
l_n(\kappa)&=&\frac { n (-1 + n + 5 \kappa) }{ (-1 + n + \kappa) (n + 4 \kappa) }, \\
p_n(\kappa)&=&\frac { n (n + \kappa) (-1 + n + 4 \kappa) (-1 + n + 5 \kappa) (-1 + 2 n + 9 \kappa) }{ 
  (-1 + n + \kappa) (-1 + n + 2 \kappa) (n + 4 \kappa)^2 (-1 + 2 n + 7 \kappa) }, \\
q_n(\kappa)&=&\frac { 2 n (n + 2 \kappa) (n + 3 \kappa) (-1 + n + 6 \kappa) (-1 + n + 8 \kappa) (-1 + 2 n + 12 \kappa) }{  (-1 + n + \kappa) (-1 + n + 3 \kappa) (n + 5 \kappa) (n + 7 \kappa) (-1 + 2 n + 7 \kappa) (2 n + 11 \kappa) } \\
r_n(\kappa)&=&\frac { n (-1 + n + 6 \kappa) }{ (-1 + n + \kappa) (n + 5 \kappa) }, \\
s_n(\kappa)&=&\frac { n (n + 3 \kappa) (-1 + n + 9 \kappa) (-1 + n + 12 \kappa) }{ 
  (-1 + n + \kappa) (-1 + n + 4 \kappa) (n + 8 \kappa) (n + 11 \kappa) }.
\eeqrn
Note that the series $z_6 P^\kappa_{n\lambda_j}$ immediately follow by duality.
\section{Conclusions}
In this paper, we have shown how to solve the Schr\"{o}dinger equation for the trigonometric Calogero-Sutherland model related to the Lie algebra $E_6$ and we have explored some properties of the energy eigenfunctions. The main point is that the use of a Weyl-invariant set of variables, the characters of the fundamental representations, leads to a formulation of the Schr\"odinger equation by means of a second order differential operator which is simple enough to make feasible a recursive method for the treatment of the spectral problem. The eigenfunctions provide a complete system of orthogonal polynomials in six variables, and these polynomials obey recurrence relations which are deformations of the Clebsch-Gordan series of the algebra. The structure of some of these recurrence relations has been fixed.
\section*{Acknowledgements} 
This work has been partially supported by the Spanish 
Ministerio de Educaci\'{o}n y Ciencia under grants BFM2003-02532 (J.F.N) and BFM2003-00936 / FISI (W.G.F and A.M.P). A.M.P would like to thank the Departament of Theoretical Physics of University of Zaragoza for hospitality.
\section*{Appendix A: Summary of results on the Lie algebra $E_6$}
In this Section, we review some standard facts about the root and weight systems of the Lie algebra $E_6$, with the aim of fixing the notation and help the reader to follow the rest of the paper. More extensive and sound treatments of these topics can be found in many excellent textbooks, see for instance \cite{ov90}, \cite{otros}. 

The complex Lie algebra $E_6$, the lowest-dimensional one in the $E$-family of exceptional Lie algebras in the Cartan-Killing classification, has dimension 78 and rank 6, as the name suggests. From the geometrical point of view, it admits (with some subtleties, see \cite{baez}) an interpretation which extends the standard-one for the classical algebras: in the same way that these correspond to the isometries of projective spaces over the first three normed division algebras ---$SO(n+1)\simeq {\rm Isom}({\bf R} P^n)$, $SU(n+1)\simeq {\rm Isom}({{\bf C}}P^n)$, $Sp(n+1)\simeq {\rm Isom}({{\bf H}}P^n)$--- $F_4$, $E_6$, $E_7$ and $E_8$ are the Lie algebras of the projective planes over extensions of the octonions, giving rise to the so-called ``magic square": $F_4\simeq {\rm Isom}({{\bf O}}P^2)$, $E_6\simeq {\rm Isom}[({{\bf C}}\otimes{{\bf O}})P^2]$,  $E_7\simeq {\rm Isom}[({{\bf H}}\otimes{{\bf O}})P^2]$, $E_8\simeq {\rm Isom}[({{\bf O}}\otimes{{\bf O}})P^2]$. In Physics, the most remarkable role played by $E_6$ is in the heterotic ten-dimensional $E_8\times E_8$ superstring theory when the extra six dimensions are compactified to a manifold of $SU(3)$ holonomy. In such a case, one of the $E_8$ breaks to an $E_6$ which gives the Grand Unification group of four-dimensional physics \cite{ramond}.  The Dynkin diagram of $E_6$, see Figure 1, 

\begin{center}
\begin{picture}(80,65)(-50,-8)
\put(0,0){\circle{8}}
\put(4,0){\line(1,0){30}}
\put(38,0){\circle{8}}
\put(42,0){\line(1,0){30}}
\put(76,0){\circle{8}}
\put(0,4){\line(0,1){30}}
\put(0,38){\circle{8}}
\put(-4,0){\line(-1,0){30}}
\put(-38,0){\circle{8}}
\put(-42,0){\line(-1,0){30}}
\put(-76,0){\circle{8}}
\put(-81,-15){$\alpha_1$}
\put(-43,-15){$\alpha_3$}
\put(-5,-15){$\alpha_4$}
\put(33,-15){$\alpha_5$}
\put(71,-15){$\alpha_6$}
\put(-4,46){$\alpha_2$}
\end{picture}
\end{center}
\bigskip
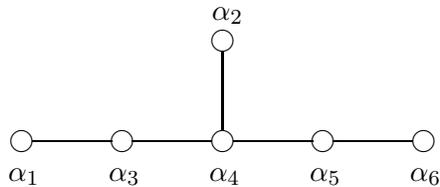
\centerline{Figure 1. The Dynkin diagram for the Lie algebra $E_6$.}
\bigskip
encodes the euclidean relations among the simple roots, which are
\beqrn
(\alpha_i,\alpha_i)&=&2,\hspace{1.5cm} i=1,2,3,4,5,6\,,\\
(\alpha_4,\alpha_i)&=&-1,\hspace{1.2cm} i=2,3,5\,,\\
(\alpha_1,\alpha_3) &=&(\alpha_5,\alpha_6) =-1\,,\\
(\alpha_i,\alpha_j)&=&0, \hspace{1.2 cm}{\rm in\ all\ other\ cases}.
\eeqrn
Therefore, the Cartan matrix reads
\bdm
A=\left(\begin{array}{cccccc}
2&0&-1&0&0&0\\
0&2&0&-1&0&0\\
-1&0&2&-1&0&0\\
0&-1&-1&2&-1&0\\
0&0&0&-1&2&-1\\
0&0&0&0&-1&2
\end{array}\right) .
\edm

We will use a realization of the simple roots in terms of a generating system $\{\varepsilon_1,\varepsilon_2,\varepsilon_3,\varepsilon_4,\varepsilon_5,\varepsilon_6,\varepsilon\}$ of ${{\bf R}}^7$ (endowed with the standard Euclidean metric) satisfying $\varepsilon_1+\varepsilon_2+\varepsilon_3+\varepsilon_4+\varepsilon_5+\varepsilon_6=0$, $(\varepsilon_i,\varepsilon_j)=-\frac{1}{6}+\delta_{ij}$, $(\varepsilon,\varepsilon)=\frac{1}{2}$ and $(\varepsilon,\varepsilon_j)=0$ \cite{ov90}. With reference to this system, we have
\begin{eqnarray}
\alpha_1&=&\varepsilon_1-\varepsilon_2,\hspace{1.5cm}\alpha_2=\varepsilon_4+\varepsilon_5+\varepsilon_6+\varepsilon\,,\nonumber\\
\alpha_3&=&\varepsilon_2-\varepsilon_3,\hspace{1.5cm}\alpha_4=\varepsilon_3-\varepsilon_4\,,\label{e61}\\
\alpha_5&=&\varepsilon_4-\varepsilon_5,\hspace{1.5cm}\alpha_6=\varepsilon_5-\varepsilon_6.\nonumber
\end{eqnarray}
The positive roots, which are given by all linear combinations of the forms
\beq
\varepsilon_i-\varepsilon_j,\ \  \varepsilon_i+\varepsilon_j+\varepsilon_k+\varepsilon, \ \ 2\varepsilon,\ \ \ \ i \neq j\neq k,\label{e62}
\eeq
can be classified by heights as indicated in the Table 1. 
\begin{table}[h]
\begin{center}
\begin{tabular}{|c|l|}
\hline
{\rm Height} & {\rm Positive roots} \\
\hline
1 & $\alpha_1,\ \alpha_2,\ \alpha_3,\ \alpha_4,\ \alpha_5,\alpha_6$ \\
\hline
2 & $\alpha_1+\alpha_3,\ \alpha_3+\alpha_4,\ \alpha_4+\alpha_5,\ \alpha_5+\alpha_6,\ \alpha_2+\alpha_4 $\\
\hline
3 & $\alpha_1+\alpha_3+\alpha_4,\ \alpha_3+\alpha_4+\alpha_5,\ \alpha_4+\alpha_5+\alpha_6,\ \alpha_2+\alpha_3+\alpha_4,$\\ & $\  \alpha_2+\alpha_4+\alpha_5$\\
\hline
4 & $\alpha_1+\alpha_3+\alpha_4+\alpha_5,\ \alpha_3+\alpha_4+\alpha_5+\alpha_6,\ \alpha_1+\alpha_2+\alpha_3+\alpha_4,$\\ & $\  \alpha_2+\alpha_3+\alpha_4+\alpha_5,\ \alpha_2+\alpha_4+\alpha_5+\alpha_6$ \\
\hline
5 & $\alpha_1+\alpha_3+\alpha_4+\alpha_5+\alpha_6,\ \alpha_1+\alpha_2+\alpha_3+\alpha_4+\alpha_5,\ \alpha_2+\alpha_3+2\alpha_4+\alpha_5,$\\ &$\ \alpha_2+\alpha_3+\alpha_4+\alpha_5+\alpha_6$ \\
\hline
6 & $ \alpha_1+\alpha_2+\alpha_3+2\alpha_4+\alpha_5,\ \alpha_1+\alpha_2+\alpha_3+\alpha_4+\alpha_5+\alpha_6,$\\ & $\ \alpha_2+\alpha_3+2 \alpha_4+\alpha_5+\alpha_6$\\
\hline
7 & $\alpha_1+\alpha_2+2\alpha_3+2\alpha_4+\alpha_5,\ \alpha_2+\alpha_3+2\alpha_4+2 \alpha_5+\alpha_6,$\\ &$\  \alpha_1+\alpha_2+\alpha_3+2\alpha_4+\alpha_5+\alpha_6$ \\
\hline
8 & $\alpha_1+\alpha_2+2 \alpha_3+2 \alpha_4+\alpha_5+\alpha_6,\ \alpha_1+\alpha_2+\alpha_3+2 \alpha_4+2 \alpha_5+\alpha_6$ \\
\hline
9 & $\alpha_1+\alpha_2+2 \alpha_3+2 \alpha_4+2 \alpha_5+\alpha_6$ \\
\hline
10 & $\alpha_1+ \alpha_2+2 \alpha_3+3 \alpha_4+2 \alpha_5+\alpha_6$\\
\hline
11 & $\alpha_1+ 2\alpha_2+2 \alpha_3+3 \alpha_4+2 \alpha_5+\alpha_6$\\
\hline
\end{tabular}
\caption[smallcaption]{Heights of positive roots.}
\end{center}
\end{table}
The fundamental weights $\lambda_k$ follow from the equation $\alpha_i=\sum_{j=1}^4 A_{ji}\lambda_j$. They are
\beqrn
\lambda_1&=&\varepsilon_1+\varepsilon\,,\\
\lambda_2&=&2\varepsilon\,,\\
\lambda_3&=&\varepsilon_1+\varepsilon_2+2\varepsilon\,,\\
\lambda_4&=&\varepsilon_1+\varepsilon_2+\varepsilon_3+3\varepsilon\,,\\
\lambda_5&=&\varepsilon_1+\varepsilon_2+\varepsilon_3+\varepsilon_4+2\varepsilon\,,\\
\lambda_6&=&\varepsilon_1+\varepsilon_2+\varepsilon_3+\varepsilon_4+\varepsilon_5+ \varepsilon .
\eeqrn
The geometry of the weight system is summarized by the relations
\beqrn
(\lambda_i,\lambda_j)=A_{ij}^{-1},
\eeqrn
with $(A_{ij}^{-1})$ being the inverse Cartan matrix.
The Weyl vector is
\bdm
\rho =\frac{1}{2}\sum_{\alpha\in {\cal R}^+}\alpha=\sum_{i=1}^6 \lambda_i=8\alpha_1+11\alpha_2+15\alpha_3+21\alpha_4+15\alpha_5+8\alpha_6\,,
\edm
with ${\cal R}^+$ being the set of positive roots of the algebra. The Weyl formula for dimensions applied to the irreducible representation associated to the integral dominant weight $\lambda=m_1\lambda_1+m_2\lambda_2+m_3\lambda_3+m_4\lambda_4+m_5\lambda_5+m_6\lambda_6$ gives
\bdm
\dim R_\lambda=\prod_{\alpha\in {\cal R}^+} \frac{(\alpha, \lambda+\rho)}{(\alpha,\rho)}=\frac{P}{2^5\cdot 3^5\cdot 4^5\cdot 5^4 \cdot 6^3 \cdot 7^3\cdot 8^2 \cdot 9\cdot 10\cdot 11}
\edm
where $P$ is a product extended to the set of positive roots in which the root $\alpha=\sum_{i=1}^6 c_i\,\alpha_i$ contributes with a factor ${\rm ht}(\alpha)+\sum_{i=1}^6 c_i\, m_i$ where ${\rm ht}(\alpha)$ is the height of $\alpha$.
In particular, for the fundamental representations, one finds:
\beqrn
\dim R_{\lambda_1}=27&\ \ \ \ \ \ \ \ \ &\dim R_{\lambda_2}=78\\
\dim R_{\lambda_3}=351&\ \ \ \ \ \ \ \ \ &\dim R_{\lambda_4}=2925\\
\dim R_{\lambda_5}=351&\ \ \ \ \ \ \ \ \ &\dim R_{\lambda_6}=27 .
\eeqrn
Note that, these dimensions reflect the fact, coming from the ${\bf Z}_2$ symmetry (duality) of the Dynkin diagram, that the representations $R_{\l_1}$ and $R_{\l_6}$ are complex conjugates. The same is true for $R_{\l_3}$ and $R_{\l_5}$, while $R_{\l_2}$ (the adjoint representation) and $R_{\l_4}$ are real.
\section*{Appendix B: Some polynomials and monomial functions}
We list here  the polynomials up to degree two, and the monomial 
functions up to degree three. Some of them are omitted for they can be obtained by duality.
\subsection*{Polynomials}
\footnotesize
\beqrn
&&P^\kappa_{200000}=z_1^2
-\frac{ 2 z_3 }{ 1 + \kappa }
-\frac{ 10 \kappa z_6 }{ (1 + \kappa) (1 + 4 \kappa) }\,,\\ 
&&P^\kappa_{110000}=z_1 z_2
-\frac{ 5 z_5 }{ 1 + 4 \kappa }
+\frac{ (6 - 95 \kappa + 24 \kappa^2) z_1 }{ (1 + 4 \kappa) (2 + 11 \kappa) }\,, \\ &&P^\kappa_{020000}=z_2^2
-\frac{ 2 z_4 }{ 1 + \kappa }
-\frac{ 8 \kappa z_1 z_6 }{ (1 + \kappa) (1 + 3 \kappa) }
+\frac{ 6 (-1 + \kappa) (1 - \kappa + 6 \kappa^2) z_2 }{ (1 + \kappa) (1 + 3 \kappa) (3 + 11 \kappa) }
+\frac{ 18 (-1 + \kappa) (2 + 13 \kappa - 7 \kappa^2 + 6 \kappa^3) }{ 
  (1 + \kappa) (1 + 3 \kappa) (2 + 11 \kappa) (3 + 11 \kappa) }\,,\\ 
 &&P^\kappa_{101000}=z_1 z_3
-\frac{ 3 z_4 }{ 1 + 2 \kappa }
+\frac{ (-2 - 35 \kappa + 10 \kappa^2) z_1 z_6 }{ (1 + 2 \kappa) (2 + 7 \kappa) }
-\frac{ 6 (-1 + \kappa) (2 + 15 \kappa) z_2 }{ (1 + 2 \kappa) (1 + 4 \kappa) (2 + 7 \kappa) }
+\frac{ 9 (-2 + 17 \kappa) }{ (1 + 2 \kappa) (1 + 4 \kappa) (2 + 7 \kappa) }\,,\\ 
&&P^\kappa_{011000}=z_2 z_3
-\frac{ 4 z_1 z_5 }{ 1 + 3 \kappa }
+\frac{ 5 (-1 + \kappa) (-2 + 3 \kappa) z_2 z_6 }{ (1 + 3 \kappa) (2 + 7 \kappa) }
-\frac{ 4 (-1 + 7 \kappa) z_1^2 }{ (1 + 3 \kappa) (1 + 5 \kappa) }\\
  &&\qquad+\frac{ 6 (-4 + 51 \kappa + 311 \kappa^2 + 41 \kappa^3 + 105 \kappa^4) z_3 }{ 
  (1 + 3 \kappa) (1 + 5 \kappa) (2 + 7 \kappa) (3 + 11 \kappa) }
+\frac{ 6 (-16 + 17 \kappa + 673 \kappa^2 - 245 \kappa^3 + 75 \kappa^4) z_6 }{ 
  (1 + 3 \kappa) (1 + 5 \kappa) (2 + 7 \kappa) (3 + 11 \kappa) }\,,\\ 
  &&P^\kappa_{002000}=z_3^2
-\frac{ 2 z_1 z_4 }{ 1 + \kappa }
-\frac{ 2 (-1 + \kappa) z_2 z_5 }{ (1 + \kappa) (1 + 2 \kappa) }
-\frac{ 8 \kappa z_1^2 z_6 }{ (1 + \kappa) (1 + 3 \kappa) }
+\frac{ 2 (-3 + 7 \kappa + 49 \kappa^2 + 37 \kappa^3 + 30 \kappa^4) z_3 z_6 }{ 
  (1 + \kappa) (1 + 2 \kappa) (1 + 3 \kappa) (3 + 7 \kappa) }\\ 
  &&\qquad- \frac{ 2 (-1 + \kappa) \kappa (17 + 69 \kappa) z_1 z_2 }{ (1 + \kappa) (1 + 2 \kappa) (1 + 3 \kappa) (3 + 7 \kappa) }
+\frac{ (-1 + \kappa) (42 + 291 \kappa + 478 \kappa^2 + 59 \kappa^3 + 150 \kappa^4) z_6^2 }{ 
  (1 + \kappa) (1 + 2 \kappa) (1 + 3 \kappa) (2 + 7 \kappa) (3 + 7 \kappa) }\\ 
  &&\qquad+\frac{ 12 (6 + 27 \kappa + 41 \kappa^2 + 103 \kappa^3 + 3 \kappa^4) z_5 }{ 
  (1 + \kappa) (1 + 2 \kappa) (1 + 3 \kappa) (2 + 7 \kappa) (3 + 7 \kappa) }
-\frac{ 6 (-2 - 7 \kappa + 58 \kappa^2 - 433 \kappa^3 + 24 \kappa^4) z_1 }{ 
  (1 + \kappa) (1 + 2 \kappa) (1 + 3 \kappa) (2 + 7 \kappa) (3 + 7 \kappa) }\,,\\ 
  &&P^\kappa_{100100}=z_1 z_4
-\frac{ 4 z_2 z_5 }{ 1 + 3 \kappa }
+\frac{ 4 (-1 + \kappa) z_1^2 z_6 }{ 1 + 5 \kappa }
-\frac{ 2 (-1 + \kappa) (5 + 21 \kappa) z_3 z_6 }{ (1 + 3 \kappa)^2 (1 + 5 \kappa) }+\frac{ (1 - \kappa) (27 + 292 \kappa + 723 \kappa^2 - 270 \kappa^3) z_1 z_2 }{ 
  (1 + 3 \kappa)^2 (1 + 5 \kappa) (3 + 10 \kappa) } \\
&&\qquad 
+\frac{ 2 (7 + 56 \kappa - 15 \kappa^2) z_6^2 }{ (1 + 3 \kappa)^2 (1 + 5 \kappa) }-\frac{ 6 (34 + 321 \kappa + 712 \kappa^2 + 55 \kappa^3 + 750 \kappa^4) z_5 }{ 
  (1 + 3 \kappa)^2 (1 + 5 \kappa) (2 + 7 \kappa) (3 + 10 \kappa) }\\
&&\qquad
+\frac{ 3 (-1 + \kappa) (42 + 703 \kappa + 1634 \kappa^2 - 2937 \kappa^3 - 270 \kappa^4) z_1 }{ 
  (1 + 3 \kappa)^2 (1 + 5 \kappa) (2 + 7 \kappa) (3 + 10 \kappa) }\,,\\ 
  &&P^\kappa_{010100}=z_2 z_4
-\frac{ 3 z_3 z_5 }{ 1 + 2 \kappa }
+\frac{ 4 (-1 + \kappa) (-1 + 2 \kappa) z_1 z_2 z_6 }{ (1 + 2 \kappa) (2 + 5 \kappa) }
+\frac{ 6 (-1 + \kappa) (-1 + 2 \kappa) (-2 + 5 \kappa) z_2^2 }{ (1 + 2 \kappa) (2 + 5 \kappa) (3 + 10 \kappa) }\\
&&\qquad-\frac{ 5 (\kappa-1) (2 + 11 \kappa) z_1 z_3 }{ (1 + 2 \kappa) (1 + 3 \kappa) (2 + 5 \kappa) }
-\frac{ 5 (\kappa-1) (2 + 11 \kappa) z_5 z_6 }{ (1 + 2 \kappa) (1 + 3 \kappa) (2 + 5 \kappa) }+\frac{ 6 (\kappa-1) (56 + 548 \kappa + 1465 \kappa^2 + 1000 \kappa^3 + 300 \kappa^4) z_4 }{ 
  (1 + 2 \kappa) (1 + 3 \kappa) (2 + 5 \kappa) (3 + 10 \kappa) (4 + 11 \kappa) }\\ 
  &&\qquad+\frac{ (-456 - 2930 \kappa + 2063 \kappa^2 + 23981 \kappa^3 - 7718 \kappa^4 + 1440 \kappa^5) z_1 z_6 }{ 
  (1 + 2 \kappa) (1 + 3 \kappa) (2 + 5 \kappa) (3 + 10 \kappa) (4 + 11 \kappa) }\\
&&\qquad+\frac{ 3 (272 + 244 \kappa - 11336 \kappa^2 - 28933 \kappa^3 + 8109 \kappa^4 - 18036 \kappa^5 + 540 \kappa^6) z_2 }{  (1 + 2 \kappa) (1 + 3 \kappa) (2 + 5 \kappa) (2 + 7 \kappa) (3 + 10 \kappa) (4 + 11 \kappa) }\\ 
   \eeqrn 
 \beqrn 
 &&\qquad+\frac{ 18 (1 - \kappa) (112 + 1200 \kappa + 2570 \kappa^2 - 1215 \kappa^3 + 1788 \kappa^4 + 180 \kappa^5) }{ 
  (1 + 2 \kappa) (1 + 3 \kappa) (2 + 5 \kappa) (2 + 7 \kappa) (3 + 10 \kappa) (4 + 11 \kappa) }\,,\\ 
  &&P^\kappa_{001100}=z_3 z_4
-\frac{ 3 z_1 z_2 z_5 }{ 1 + 2 \kappa }
-\frac{ 5 (-1 + \kappa) z_2^2 z_6 }{ (1 + 2 \kappa) (1 + 3 \kappa) }
+\frac{ 4 (-1 + \kappa) (-1 + 2 \kappa) z_1 z_3 z_6 }{ (1 + 2 \kappa) (2 + 5 \kappa) }
-\frac{ 5 (-1 + \kappa) z_5^2 }{ (1 + 2 \kappa) (1 + 3 \kappa) }\\ 
&&\qquad+\frac{ (-42 + 25 \kappa + 444 \kappa^2 + 263 \kappa^3 + 150 \kappa^4) z_4 z_6 }{ 
  (1 + 2 \kappa) (1 + 3 \kappa) (2 + 5 \kappa) (3 + 7 \kappa) }
-\frac{ 5 (-1 + \kappa) (2 + 11 \kappa) z_1^2 z_2 }{ (1 + 2 \kappa) (1 + 3 \kappa) (2 + 5 \kappa) }\\ 
&&\qquad+\frac{ 4 (-15 - 32 \kappa + 276 \kappa^2 + 626 \kappa^3 - 105 \kappa^4 + 90 \kappa^5) z_1 z_6^2 }{ 
  (1 + 2 \kappa) (1 + 3 \kappa)^2 (2 + 5 \kappa) (3 + 7 \kappa) }
+\frac{ 2 (-1 + \kappa) (84 + 574 \kappa + 904 \kappa^2 + 69 \kappa^3 + 315 \kappa^4) z_2 z_3 }{ 
  (1 + 2 \kappa) (1 + 3 \kappa) (2 + 5 \kappa)^2 (3 + 7 \kappa) }\\ 
  &&\qquad\frac{ 4 (1-\kappa) (30 + 109 \kappa - 87 \kappa^2 + 158 \kappa^3 + 1680 \kappa^4) z_1 z_5 }{(1 + 2 \kappa) (1 + 3 \kappa)^2 (2 + 5 \kappa)^2 (3 + 7 \kappa) }+\frac{ ( \kappa-1) (-36 - 1080 \kappa - 8095 \kappa^2 - 12988 \kappa^3 + 5847 \kappa^4) z_1^2 }{ 
  (1 + 2 \kappa) (1 + 3 \kappa)^2 (2 + 5 \kappa)^2 (3 + 7 \kappa) }\\ 
  &&\qquad+\frac{ 2 (-1 + \kappa) (12 + 734 \kappa + 5474 \kappa^2 + 9705 \kappa^3 - 1620 \kappa^4 + 675 \kappa^5) z_2 z_6 }{ 
  (1 + 2 \kappa) (1 + 3 \kappa)^2 (2 + 5 \kappa)^2 (3 + 7 \kappa) }\\ 
  &&\qquad+\frac{ 3 (1 -\kappa) (1 + 6 \kappa) (44 + 492 \kappa - 41 \kappa^2 + 252 \kappa^3 + 45 \kappa^4) z_3 }{ 
  (1 + 2 \kappa) (1 + 3 \kappa)^2 (2 + 5 \kappa)^2 (3 + 7 \kappa) }\\ 
&&\qquad-\frac{ 3 (-120 - 3020 \kappa - 14654 \kappa^2 + 9383 \kappa^3 + 99779 \kappa^4 - 34713 \kappa^5 + 12555 \kappa^6 + 
     1350 \kappa^7) z_6 }{ (1 + 2 \kappa) (1 + 3 \kappa)^2 (2 + 5 \kappa)^2 (2 + 7 \kappa) (3 + 7 \kappa) }\,,\\ 
     &&P^\kappa_{000200}=z_4^2
-\frac{ 2 z_2 z_3 z_5 }{ 1 + \kappa }
-\frac{ 2 (-1 + \kappa) z_3^2 z_6 }{ (1 + \kappa) (1 + 2 \kappa) }
-\frac{ 2 (-1 + \kappa) z_1 z_5^2 }{ (1 + \kappa) (1 + 2 \kappa) }
-\frac{ 2 (-1 + \kappa) z_1 z_2^2 z_6 }{ (1 + \kappa) (1 + 2 \kappa) }
-\frac{ 2 (-1 + \kappa) (-1 + 2 \kappa) z_2^3 }{ (1 + \kappa) (1 + 2 \kappa) (1 + 3 \kappa) }\\ 
&&\qquad+\frac{ 4 (-3 + 5 \kappa + 6 \kappa^2 + 4 \kappa^3) z_1 z_4 z_6 }{ (1 + \kappa) (1 + 2 \kappa) (3 + 5 \kappa) }
-\frac{ 2 (-1 + \kappa) (-3 + 2 \kappa + 28 \kappa^2) z_2 z_5 z_6 }{ (1 + \kappa) (1 + 2 \kappa)^2 (3 + 5 \kappa)  }
-\frac{ 2 (-1 + \kappa) (-3 + 2 \kappa + 28 \kappa^2) z_1 z_2 z_3 }{ (1 + \kappa) (1 + 2 \kappa)^2 (3 + 5 \kappa)  }\\ 
&&\qquad+\frac{ 6 (-1 + \kappa) (-15 + 2 \kappa + 335 \kappa^2 + 754 \kappa^3 + 436 \kappa^4 + 120 \kappa^5) z_2 z_4 }{ 
  5 (1 + \kappa) (1 + 2 \kappa)^3 (1 + 3 \kappa) (3 + 5 \kappa) }
+\frac{ 16 (-1 + \kappa) (3 + 10 \kappa + 3 \kappa^2 + 2 \kappa^3) z_1^2 z_6^2 }{ 
  (1 + \kappa) (1 + 2 \kappa) (2 + 5 \kappa) (3 + 5 \kappa) }\\
  &&\qquad-\frac{ 4 (-18 - 65 \kappa - 82 \kappa^2 - 109 \kappa^3 + 22 \kappa^4) z_3 z_6^2 }{ 
  (1 + \kappa) (1 + 2 \kappa)^2 (2 + 5 \kappa) (3 + 5 \kappa) }
-\frac{ 4 (-18 - 65 \kappa - 82 \kappa^2 - 109 \kappa^3 + 22 \kappa^4) z_1^2 z_5 }{ 
  (1 + \kappa) (1 + 2 \kappa)^2 (2 + 5 \kappa) (3 + 5 \kappa) }\\
  &&\qquad-\frac{ 4 (150 + 1507 \kappa + 6668 \kappa^2 + 17329 \kappa^3 + 27482 \kappa^4 + 23584 \kappa^5 + 9800 \kappa^6 + 
     4200 \kappa^7) z_3 z_5 }{ 5 (1 + \kappa) (1 + 2 \kappa)^4 (1 + 3 \kappa) (2 + 5 \kappa) (3 + 5 \kappa) }\\
     &&\qquad-\frac{ 2 (-1 + \kappa) (6 + 39 \kappa - 118 \kappa^2 - 453 \kappa^3 + 70 \kappa^4) z_6^3 }{ 
  (1 + \kappa) (1 + 2 \kappa)^2 (1 + 3 \kappa) (2 + 5 \kappa) (3 + 5 \kappa) }\\
  &&\qquad+\frac{ 2 (-1 + \kappa) (-30 - 21 \kappa + 3383 \kappa^2 + 22456 \kappa^3 + 52408 \kappa^4 + 39680 \kappa^5 - 
     3216 \kappa^6 + 2880 \kappa^7) z_1 z_2 z_6 }{ 
  5 (1 + \kappa) (1 + 2 \kappa)^4 (1 + 3 \kappa) (2 + 5 \kappa) (3 + 5 \kappa) }\\
  &&\qquad-\frac{ 4 (-1 + \kappa) (-6 - 37 \kappa + 225 \kappa^2 + 1328 \kappa^3 + 1224 \kappa^4 - 616 \kappa^5 + 600 \kappa^6) z_5 
   z_6 }{ (1 + \kappa) (1 + 2 \kappa)^4 (1 + 3 \kappa) (2 + 5 \kappa) (3 + 5 \kappa) }\\
  &&\qquad-\frac{ 2 (-1 + \kappa) (6 + 39 \kappa - 118 \kappa^2 - 453 \kappa^3 + 70 \kappa^4) z_1^3 }{ 
  (1 + \kappa) (1 + 2 \kappa)^2 (1 + 3 \kappa) (2 + 5 \kappa) (3 + 5 \kappa) }\\
  &&\qquad+\frac{ 9 (-1 + \kappa) (-60 - 784 \kappa - 4813 \kappa^2 - 15896 \kappa^3 - 24883 \kappa^4 - 9500 \kappa^5 + 
     9296 \kappa^6 + 80 \kappa^7 + 1200 \kappa^8) z_2^2 }{ 
  5 (1 + \kappa) (1 + 2 \kappa)^4 (1 + 3 \kappa) (2 + 5 \kappa)^2 (3 + 5 \kappa) }\\
  &&\qquad-\frac{ 4 (-1 + \kappa) (-6 - 37 \kappa + 225 \kappa^2 + 1328 \kappa^3 + 1224 \kappa^4 - 616 \kappa^5 + 600 \kappa^6) z_1 
   z_3 }{ (1 + \kappa) (1 + 2 \kappa)^4 (1 + 3 \kappa) (2 + 5 \kappa) (3 + 5 \kappa) }
 -\frac{Az_1z_6}{a}-\frac{Bz_2}{a}-\frac{Cz_4}{a}+\frac{D}{a}\,,\\
  &&P^\kappa_{001010}=z_3 z_5
-\frac{ 4 z_1 z_2 z_6 }{ 1 + 3 \kappa }
-\frac{ 9 (-1 + \kappa) z_2^2 }{ (1 + 3 \kappa) (1 + 4 \kappa) }
+\frac{ 5 (-1 + \kappa) (-2 + 3 \kappa) z_1 z_3 }{ (1 + 3 \kappa) (2 + 7 \kappa) }
+\frac{ 5 (-1 + \kappa) (-2 + 3 \kappa) z_5 z_6 }{ (1 + 3 \kappa) (2 + 7 \kappa) }\\
&&\qquad+\frac{ 24 (-1 + \kappa + 21 \kappa^2 + 9 \kappa^3) z_4 }{ (1 + 3 \kappa)^2 (1 + 4 \kappa) (2 + 7 \kappa) }+
\frac{ (-44 + 140 \kappa + 3413 \kappa^2 + 7150 \kappa^3 - 5079 \kappa^4 + 900 \kappa^5) z_1 z_6 }{ 
  (1 + 3 \kappa)^2 (1 + 4 \kappa) (2 + 7 \kappa)^2 }\\
&&\qquad-\frac{ 36 \kappa (16 + 26 \kappa - 231 \kappa^2 + 9 \kappa^3) z_2 }{ (1 + 3 \kappa)^2 (1 + 4 \kappa) (2 + 7 \kappa)^2 }-\frac{ 108 \kappa (6 + 103 \kappa + 311 \kappa^2 - 123 \kappa^3 + 63 \kappa^4) }{ 
  (1 + 3 \kappa)^2 (1 + 4 \kappa) (1 + 5 \kappa) (2 + 7 \kappa)^2 }\,,\\
  &&P^\kappa_{100001}=z_1 z_6
-\frac{ 6 z_2 }{ 1 + 5 \kappa }-\frac{9(-1+7\kappa)}{(1+5\kappa)(1+8\kappa)}\,,
\eeqrn
where the coefficients $A, B, C, D$ and $a$ are
\beqrn
&&A=8 \kappa (-12 - 6872 \kappa - 74937 \kappa^2 - 237510 \kappa^3 - 15495 \kappa^4 + 979026 \kappa^5 + 989844 \kappa^6 - 199504 \kappa^7 + 142260 \kappa^8 + 10800 \kappa^9)\,,\\
&&B=18 (1-\kappa) (180 +2196 \kappa +12403 \kappa^2 + 34729 \kappa^3 + 9833\kappa^4 - 153277 \kappa^5 - 225096 \kappa^6 - 37608 \kappa^7 - 36240 \kappa^8 - 3600 \kappa^9)\,,\\
\eeqrn
\beqrn
&&C= 6 (420 + 6424 \kappa + 50807 \kappa^2 + 228922 \kappa^3 + 594476 \kappa^4 + 938974 \kappa^5 + 
     1027217 \kappa^6 + 835680 \kappa^7 + 400680 \kappa^8 + 132000 \kappa^9 \\
     &&\qquad+ 18000 \kappa^{10})\,,\\  
&&D=27 (120 + 1772 \kappa + 7970 \kappa^2 + 5421 \kappa^3 + 21440 \kappa^4 + 503710 \kappa^5 + 1712910 \kappa^6 + 
  1652129 \kappa^7 + 44920 \kappa^8 + 259768 \kappa^9\\
  &&\qquad + 19840 \kappa^{10} + 3600 \kappa^{11})\,, \\
  &&a=  5 (1 + \kappa) (1 + 2 \kappa)^4 (1 + 3 \kappa) (2 + 5 \kappa)^2 (3 + 5 \kappa) (3 + 7 \kappa)\,.
\eeqrn
\normalsize
\subsection*{Monomial functions}
\footnotesize
\beqrn
&&M_{200000}=  z_1^2-2 z_3\,,\\
&&M_{110000}=z_1 z_2-5 z_5+3 z_1\,,\\
&&M_{020000}=  z_2^2-2 z_4-2 z_2-6\,,\\
&&M_{101000}=  z_1 z_3-3 z_4-z_1 z_6+6 z_2-9\,,\\
&&M_{011000}=  z_2 z_3-4 z_1 z_5+5 z_2 z_6+4 z_1^2-4 z_3-16z_6\,,\\
&&M_{002000}=  z_3^2-2 z_1 z_4+2 z_2 z_5-2 z_3 z_6-7 z_6^2+12 z_5+2 z_1 \,,\\ 
&&M_{100100}=  z_1 z_4-4 z_2 z_5-4 z_1^2 z_6+10 z_3 z_6+9 z_1 z_2+14 z_6^2
-34 z_5-21 z_1\,,\\
&&M_{010100}=  z_2 z_4-3 z_3 z_5+2 z_1 z_2 z_6-2 z_2^2+5 z_1 z_3+5 z_5 z_6
-14 z_4-19 z_1 z_6+17 z_2+42\,,\\
&&M_{001100}=  z_3 z_4
-3 z_1 z_2 z_5
+5 z_2^2 z_6
+2 z_1 z_3 z_6
+5 z_5^2
-7 z_4 z_6
+5 z_1^2 z_2
-10 z_1 z_6^2
-14 z_2 z_3
+10 z_1 z_5
-3 z_1^2
-2 z_2 z_6\\
&&\qquad+11 z_3
+15 z_6\,,\\
&&M_{000200}=  z_4^2
-2 z_2 z_3 z_5
+2 z_3^2 z_6
+2 z_1 z_5^2
+2 z_1 z_2^2 z_6
-2 z_2^3
-4 z_1 z_4 z_6
-2 z_2 z_5 z_6
-2 z_1 z_2 z_3
+6 z_2 z_4
-8 z_1^2 z_6^2
+12 z_3 z_6^2\\
&&\qquad+12 z_1^2 z_5
-20 z_3 z_5
+2 z_6^3
+2 z_1 z_2 z_6
-4 z_5 z_6
+2 z_1^3
+9 z_2^2
-4 z_1 z_3
-14 z_4
-18 z_2+9\,,\\
&&M_{001010}=  z_3 z_5
-4 z_1 z_2 z_6
+9 z_2^2
+5 z_1 z_3
+5 z_5 z_6
-12 z_4
-11 z_1 z_6\,,\\
&&M_{100001}=  z_1 z_6
-6 z_2+9\,,\\
&&M_{300000}=  z_1^3-3 z_1 z_3+3 z_4\,,\\
&&M_{210000}=  z_1^2 z_2
-2 z_2 z_3
-z_1 z_5
-z_1^2
+5 z_2 z_6
-5 z_3-9 z_6\,,\\ 
&&M_{120000}=  z_1 z_2^2
-2 z_1 z_4
-z_2 z_5
+4 z_3 z_6
-4 z_6^2
-8 z_1 z_2
+9 z_5
+7 z_1\,,\\
&&M_{030000}=  z_2^3
-3 z_2 z_4
+3 z_3 z_5
-3 z_1 z_2 z_6
+3 z_4
+3 z_1 z_6-9\,,\\
&&M_{201000}=  z_1^2 z_3
-2 z_3^2
-z_1 z_4
-z_1^2 z_6
+4 z_2 z_5
-2 z_3 z_6
-4 z_6^2
+z_1 z_2
-z_5
+8 z_1\,,\\
&&M_{111000}=  z_1 z_2 z_3
-3 z_2 z_4
-4 z_1^2 z_5
+6 z_3 z_5
+7 z_1 z_2 z_6
-10 z_5 z_6
-12 z_2^2
+4 z_1^3
-19 z_1 z_3
+33 z_4
+7 z_1 z_6
+15 z_2-9\,,\\ 
&&M_{021000}=  z_2^2 z_3
-2 z_3 z_4
-z_1 z_2 z_5
+5 z_5^2
+4 z_1 z_3 z_6
-8 z_4 z_6
-6 z_1^2 z_2
-12 z_1 z_6^2
+3 z_2 z_3
+11 z_1 z_5
+14 z_1^2
+18 z_2 z_6\\
&&\qquad-18 z_3-14 z_6\,,\\
&&M_{102000}=  z_1 z_3^2
-2 z_1^2 z_4
-z_3 z_4
+5 z_1 z_2 z_5
-5 z_5^2
-5 z_2^2 z_6
-5 z_1 z_3 z_6
+10 z_4 z_6
+5 z_2 z_3
+4 z_1 z_6^2
-3 z_1 z_5
+6 z_2 z_6\\
&&\qquad-10 z_3-z_6\,,\\
&&M_{012000}=  z_2 z_3^2
-2 z_1 z_2 z_4
-z_1 z_3 z_5
+2 z_2^2 z_5
+3 z_4 z_5
+4 z_1^2 z_2 z_6
-9 z_2 z_3 z_6
-6 z_1^2 z_3
-3 z_1 z_5 z_6
-4 z_1 z_2^2
+11 z_3^2
+8 z_1 z_4\\
&&\qquad
+2 z_1^2 z_6-2 z_2 z_6^2
+6 z_2 z_5
+11 z_3 z_6
-12 z_1 z_2
-6 z_6^2
+19 z_5
+18 z_1\,,\\
&&M_{003000}=  z_3^3
-3 z_1 z_3 z_4
+3 z_4^2
+3 z_1^2 z_2 z_5
-3 z_2 z_3 z_5
-3 z_1 z_2^2 z_6
-3 z_1^2 z_3 z_6
+3 z_3^2 z_6
-3 z_1 z_5^2
+6 z_1 z_4 z_6
+3 z_1^2 z_6^2
+3 z_2 z_5 z_6\\
&&\qquad
+3 z_2^3+3 z_1 z_2 z_3
-9 z_2 z_4
-3 z_1^2 z_5
-8 z_6^3
-3 z_1 z_2 z_6
+21 z_5 z_6
-21 z_4\,,\\
&&M_{200100}=  z_1^2 z_4
-2 z_3 z_4
-z_1 z_2 z_5
-4 z_1^3 z_6
+5 z_5^2
+5 z_2^2 z_6
+12 z_1 z_3 z_6
-19 z_4 z_6
+4 z_1^2 z_2
-11 z_2 z_3
-8 z_1 z_6^2
+8 z_1^2
+4 z_2 z_6\\
&&\qquad-6 z_3-7 z_6\,,\\
&&M_{110100}=  z_1 z_2 z_4
-3 z_1 z_3 z_5
-4 z_2^2 z_5
+6 z_4 z_5
+2 z_1^2 z_2 z_6
+7 z_2 z_3 z_6
+5 z_1^2 z_3
-7 z_1 z_2^2
-3 z_1 z_5 z_6
-15 z_3^2
+2 z_1 z_4
-3 z_1^2 z_6\\ 
&&\qquad-z_2 z_6^2
+15 z_2 z_5
-6 z_3 z_6
+24 z_1 z_2
+9 z_6^2
-16 z_5
-28 z_1\,,\\
&&M_{020100}=  z_2^2 z_4
-2 z_4^2
-z_2 z_3 z_5
+4 z_3^2 z_6
+4 z_1 z_5^2
-6 z_1 z_4 z_6
-3 z_2 z_5 z_6
-3 z_1 z_2 z_3
-8 z_1^2 z_6^2
+4 z_3 z_6^2
+6 z_2 z_4
+4 z_1^2 z_5\\
&&\qquad
-8 z_3 z_5+8 z_6^3
+27 z_1 z_2 z_6
+8 z_1^3
-22 z_5 z_6
-19 z_2^2
-22 z_1 z_3
+20 z_4
-14 z_1 z_6
-4 z_2
+42\,,\\
&&M_{101100}=  z_1 z_3 z_4
-3 z_4^2
-3 z_1^2 z_2 z_5
+4 z_2 z_3 z_5
+2 z_1^2 z_3 z_6
+7 z_1 z_2^2 z_6
-4 z_3^2 z_6
+7 z_1 z_5^2
-9 z_1 z_4 z_6
-10 z_1^2 z_6^2
-20 z_2 z_5 z_6\\
&&\qquad-12 z_2^3
+5 z_1^3 z_2
-20 z_1 z_2 z_3
+12 z_3 z_6^2
+45 z_2 z_4
+2 z_1^2 z_5
+40 z_6^3
+12 z_3 z_5
+24 z_1 z_2 z_6
+3 z_1^3
-92 z_5 z_6
-21 z_1 z_3
-18 z_2^2\\
&&\qquad+96 z_4
-7 z_1 z_6
+33 z_2
-9\,,\\
&&M_{011100}=  z_2 z_3 z_4
-3 z_3^2 z_5
-3 z_1 z_2^2 z_5
+5 z_2^3 z_6
+4 z_1 z_4 z_5
+8 z_1 z_2 z_3 z_6
+7 z_2 z_5^2
-22 z_2 z_4 z_6
-5 z_1 z_3^2
-4 z_1^2 z_5 z_6
-5 z_1^2 z_2^2\\
\eeqrn
\beqrn
&&\qquad+z_3 z_5 z_6
-10 z_1 z_2 z_6^2
+8 z_5 z_6^2
+8 z_1^2 z_4
-3 z_2^2 z_3
+8 z_3 z_4
+19 z_1 z_2 z_5
-2 z_2^2 z_6
+4 z_1^3 z_6
-23 z_5^2
+z_1 z_3 z_6
-4 z_1^2 z_2
+8 z_4 z_6\\
&&\qquad+2 z_1 z_6^2
+10 z_2 z_3
-18 z_1 z_5
+17 z_1^2
+3 z_2 z_6
-14 z_3
-6 z_6\,,\\
&&M_{002100}=  z_3^2 z_4
-2 z_1 z_4^2
-z_1 z_2 z_3 z_5
+5 z_2 z_4 z_5
+4 z_1^2 z_5^2
+4 z_1^2 z_2^2 z_6
-7 z_3 z_5^2
-6 z_1^2 z_4 z_6
-7 z_2^2 z_3 z_6
-3 z_1^2 z_2 z_3
+10 z_3 z_4 z_6\\
&&\qquad
-4 z_1 z_2^3-9 z_1 z_2 z_5 z_6
+6 z_2 z_3^2
-8 z_1^3 z_6^2
+5 z_5^2 z_6
+5 z_2^2 z_6^2
+9 z_1 z_2 z_4
+24 z_1 z_3 z_6^2
+13 z_2^2 z_5
+4 z_1^3 z_5
-8 z_1 z_3 z_5
-8 z_4 z_6^2\\
&&\qquad-4 z_1 z_6^3
-18 z_4 z_5
+7 z_1^2 z_2 z_6
-23 z_2 z_3 z_6
+4 z_1 z_5 z_6
+8 z_1^4
-29 z_1^2 z_3
-20 z_2 z_6^2
+4 z_1 z_2^2
+16 z_3^2
+z_1 z_4\\
&&\qquad+z_1^2 z_6
+10 z_2 z_5
+24 z_3 z_6
-z_1 z_2
+19 z_6^2
-16 z_5
-27 z_1\,,\\
&&M_{100200}=  z_1 z_4^2
-2 z_1 z_2 z_3 z_5
-z_2 z_4 z_5
+2 z_1^2 z_5^2
+2 z_1 z_3^2 z_6
+2 z_1^2 z_2^2 z_6
+3 z_3 z_5^2
-4 z_1^2 z_4 z_6
+3 z_2^2 z_3 z_6
-2 z_1^2 z_2 z_3
-5 z_3 z_4 z_6\\
&&\qquad
-7 z_1 z_2^3-7 z_1 z_2 z_5 z_6
-3 z_2 z_3^2
-8 z_1^3 z_6^2
+21 z_1 z_2 z_4
+14 z_1 z_3 z_6^2
+7 z_2^2 z_5
+12 z_1^3 z_5
-25 z_1 z_3 z_5
-z_4 z_6^2
+8 z_1 z_6^3
+3 z_4 z_5\\
&&\qquad+13 z_1^2 z_2 z_6
-19 z_2 z_3 z_6
-17 z_1 z_5 z_6
+2 z_1^4
-16 z_1^2 z_3
-8 z_2 z_6^2
+10 z_1 z_2^2
+27 z_3^2
-11 z_1 z_4
-z_1^2 z_6
+7 z_2 z_5\\
&&\qquad+11 z_3 z_6
-31 z_1 z_2
-8 z_6^2
+37 z_5
+22 z_1\,,\\
&&M_{010200}=  z_2 z_4^2
-2 z_2^2 z_3 z_5
-z_3 z_4 z_5
+5 z_2 z_3^2 z_6
+5 z_1 z_2 z_5^2
-5 z_3^3
-5 z_5^3
+2 z_1 z_2^3 z_6
-9 z_1 z_2 z_4 z_6
-5 z_2^2 z_5 z_6\\
&&\qquad-5 z_1 z_3 z_5 z_6
+15 z_4 z_5 z_6
-5 z_1 z_2^2 z_3
-2 z_2^4
+15 z_1 z_3 z_4
-6 z_1^2 z_2 z_6^2
+7 z_2 z_3 z_6^2
+14 z_2^2 z_4
-16 z_4^2
+11 z_1 z_5 z_6^2\\
&&\qquad+7 z_1^2 z_2 z_5
-10 z_2 z_3 z_5
+11 z_1^2 z_3 z_6
+4 z_1 z_2^2 z_6
-12 z_1 z_5^2
-12 z_3^2 z_6
-17 z_1 z_4 z_6
-5 z_1^2 z_6^2
-10 z_1^3 z_2
-10 z_2 z_6^3\\
&&\qquad+29 z_2 z_5 z_6
+29 z_1 z_2 z_3
+3 z_3 z_6^2
+10 z_6^3
-21 z_2 z_4
+3 z_1^2 z_5
-34 z_3 z_5
+7 z_1 z_2 z_6
-14 z_5 z_6
+10 z_1^3
-14 z_1 z_3\\
&&\qquad-6 z_2^2
-6 z_4
-34 z_1 z_6
+38 z_2
+42\,,\\
&&M_{001200}=  z_3 z_4^2
-2 z_2 z_3^2 z_5
+2 z_3^3 z_6
-z_1 z_2 z_4 z_5
+5 z_1 z_3 z_5^2
+4 z_2^2 z_5^2
-7 z_4 z_5^2
+5 z_1 z_2^2 z_3 z_6
-7 z_2^2 z_4 z_6
-9 z_1 z_3 z_4 z_6\\
&&\qquad+14 z_4^2 z_6
-5 z_1^2 z_2 z_5 z_6
-6 z_1^2 z_3 z_6^2
-5 z_1^2 z_2^3
-6 z_2 z_3 z_5 z_6
-5 z_1 z_2 z_3^2
+15 z_1^2 z_2 z_4
+6 z_3^2 z_6^2
+7 z_1^2 z_3 z_5
+2 z_2^3 z_3\\
&&\qquad+z_2 z_3 z_4
-4 z_3^2 z_5
+24 z_1 z_4 z_6^2
+8 z_1 z_2^2 z_5
-12 z_2 z_5 z_6^2
-33 z_1 z_4 z_5
+2 z_1^2 z_6^3
+11 z_1^3 z_2 z_6
+8 z_3 z_6^3
+8 z_6^4
-28 z_1 z_2 z_3 z_6\\
&&\qquad+16 z_2 z_5^2
+7 z_1^2 z_2^2
+2 z_2 z_4 z_6
-z_1^2 z_5 z_6
-6 z_3 z_5 z_6
+11 z_2^2 z_3
-10 z_1^3 z_3
+38 z_1 z_3^2
-29 z_1^2 z_4
-22 z_3 z_4
-z_1^3 z_6\\
&&\qquad-z_1 z_2 z_5
-20 z_5 z_6^2
-9 z_1 z_3 z_6
-20 z_1^2 z_2
-5 z_2^2 z_6
+22 z_5^2
-3 z_4 z_6
+14 z_2 z_3
-24 z_1 z_6^2\\
&&\qquad+29 z_1 z_5
+10 z_2 z_6
+27 z_3
+17 z_6\,,\\
&&M_{000300}=  z_4^3
-3 z_2 z_3 z_4 z_5
+3 z_3^2 z_5^2
+3 z_1 z_2^2 z_5^2
+3 z_2^2 z_3^2 z_6
-3 z_3^2 z_4 z_6
-3 z_1 z_2^2 z_4 z_6
-3 z_1 z_4 z_5^2
-3 z_2 z_3^3
-3 z_2^3 z_5 z_6
+6 z_1 z_4^2 z_6\\
&&\qquad-9 z_1 z_2 z_3 z_5 z_6
-3 z_2 z_5^3
+12 z_2 z_4 z_5 z_6
-3 z_1 z_2^3 z_3
+3 z_3 z_5^2 z_6
+9 z_1^2 z_4 z_6^2
+12 z_1 z_2 z_3 z_4
+3 z_2^2 z_3 z_6^2
+6 z_2^3 z_4
-18 z_2 z_4^2\\
&&\qquad+3 z_1^2 z_2^2 z_5
-12 z_3 z_4 z_6^2
+3 z_1 z_3^2 z_5
-12 z_1^2 z_4 z_5
+3 z_2^2 z_6^3
-4 z_1^3 z_6^3
+9 z_3 z_4 z_5
+3 z_1^2 z_3^2
+3 z_5^2 z_6^2
-12 z_1 z_2 z_4 z_6\\
&&\qquad+12 z_1 z_3 z_6^3
-24 z_4 z_6^3
-3 z_5^3
+12 z_1^3 z_5 z_6
-33 z_1 z_3 z_5 z_6
-3 z_3^3
+3 z_1^3 z_2^2
-24 z_1^3 z_4
+54 z_4 z_5 z_6
+54 z_1 z_3 z_4
-45 z_4^2\\
&&\qquad-9 z_2 z_3 z_5
-9 z_1 z_2^2 z_6
-6 z_1 z_5 z_6^2
-6 z_2 z_6^3
-6 z_1^2 z_3 z_6
-6 z_1^3 z_2
+27 z_1 z_4 z_6
+9 z_2 z_5 z_6
+12 z_2^3
+3 z_1^2 z_6^2\\
&&\qquad+9 z_1 z_2 z_3
+9 z_1^2 z_5
-36 z_2 z_4
+9 z_3 z_6^2
+9 z_3 z_5
-18 z_1 z_2 z_6
+3 z_6^3
+3 z_1^3
+9 z_4
+9 z_1 z_6
-9\,,\\
&&M_{101010}=  z_1 z_3 z_5
-3 z_4 z_5
-4 z_1^2 z_2 z_6
+5 z_1^2 z_3
+6 z_2 z_3 z_6
+9 z_1 z_2^2
+7 z_1 z_5 z_6
-10 z_3^2
-13 z_1 z_4
-13 z_1^2 z_6
-10 z_2 z_6^2
-7 z_2 z_5\\
&&\qquad+16 z_3 z_6
+50 z_6^2
+11 z_1 z_2
-57 z_5
-59 z_1\,,\\
&&M_{011010}=  z_2 z_3 z_5
-4 z_3^2 z_6
-4 z_1 z_5^2
-4 z_1 z_2^2 z_6
+9 z_2^3
+12 z_1 z_4 z_6
+11 z_2 z_5 z_6
+11 z_1 z_2 z_3
-45 z_2 z_4
+16 z_1^2 z_6^2
-28 z_3 z_6^2\\
&&\qquad
-28 z_1^2 z_5+45 z_3 z_5
-16 z_6^3
-17 z_1 z_2 z_6
+45 z_5 z_6
-16 z_1^3
+45 z_1 z_3
-27 z_4
-3 z_1 z_6
+18 z_2
-27\,,\\
&&M_{002010}=  z_3^2 z_5
-2 z_1 z_4 z_5
+2 z_2 z_5^2
-z_1 z_2 z_3 z_6
+3 z_2 z_4 z_6
+4 z_1^2 z_5 z_6
+5 z_1^2 z_2^2
-9 z_3 z_5 z_6
-3 z_1 z_2 z_6^2
-8 z_1^2 z_4\\
&&\qquad-9 z_2^2 z_3
-2 z_5 z_6^2
+16 z_3 z_4
-6 z_1 z_2 z_5
+7 z_5^2
-12 z_1^3 z_6
+8 z_2^2 z_6
+37 z_1 z_3 z_6
-9 z_4 z_6
+2 z_1^2 z_2
+z_1 z_6^2\\
&&\qquad-27 z_2 z_3
+6 z_1 z_5
+13 z_1^2
-11 z_2 z_6
-14 z_3
+13 z_6\,,\\
&&M_{001110}=  z_3 z_4 z_5
-3 z_1 z_2 z_5^2
-3 z_2 z_3^2 z_6
+5 z_3^3
+4 z_1 z_2 z_4 z_6
+8 z_1 z_3 z_5 z_6
+5 z_5^3
+7 z_2^2 z_5 z_6
+7 z_1 z_2^2 z_3
-22 z_4 z_5 z_6\\
&&\qquad
-4 z_1^2 z_2 z_6^2
-18 z_2^2 z_4+z_2 z_3 z_6^2
-22 z_1 z_3 z_4
+36 z_4^2
+z_1^2 z_2 z_5
-10 z_1 z_5 z_6^2
+8 z_2 z_6^3
-8 z_2 z_3 z_5
-10 z_1^2 z_3 z_6
+8 z_3^2 z_6\\
&&\qquad
+8 z_1 z_5^2
+9 z_2^3+52 z_1 z_4 z_6
+16 z_1^2 z_6^2
-40 z_2 z_5 z_6
+8 z_1^3 z_2
-40 z_1 z_2 z_3
-z_3 z_6^2
-z_1^2 z_5
+9 z_2 z_4
+24 z_3 z_5\\
&&\qquad
-23 z_1 z_2 z
+18 z_2^2+27 z_1 z_3
+27 z_5 z_6
-63 z_4
-9 z_1 z_6
-36 z_2
-27\,,\\
&&M_{200001}=  z_1^2 z_6
-2 z_3 z_6
-z_1 z_2
+5 z_5
-4 z_1\,,\\
&&M_{110001}=  z_1 z_2 z_6
-6 z_2^2
-5 z_1 z_3
-5 z_5 z_6
+21 z_4
+24 z_1 z_6
-30 z_2
-9\,,\\
&&M_{101001}=  z_1 z_3 z_6
-3 z_4 z_6
-z_1 z_6^2
-5 z_1^2 z_2
+8 z_2 z_3
+9 z_1 z_5
+z_1^2
-9 z_2 z_6
+3 z_3
+2 z_6\,,\\
&&M_{100101}=  z_1 z_4 z_6
-4 z_2 z_5 z_6
-4 z_1 z_2 z_3
-4 z_1^2 z_6^2
+9 z_2 z_4
+10 z_3 z_6^2
+10 z_1^2 z_5
-9 z_3 z_5
-z_1 z_2 z_6
+14 z_6^3
-39 z_5 z_6\\
&&\qquad+14 z_1^3
-39 z_1 z_3
+27 z_4
+24 z_1 z_6
-81.
\eeqrn
\normalsize
\section*{Appendix C: Deformed quadratic Clebsch-Gordan series}
\footnotesize
\beqrn
&&P^\kappa_{100000}\times P^\kappa_{100000}=
  P^\kappa_{ 200000}
+\frac{ 2 }{ 1 + \kappa }  P^\kappa_{ 001000}
+\frac{ 10 (1 + 3 \kappa) }{ (1 + 4 \kappa) (1 + 7 \kappa) } P^\kappa_{ 000001}\,,\\
&&P^\kappa_{100000}\times P^\kappa_{010000}=
  P^\kappa_{ 110000}
+\frac{ 5 }{ 1 + 4 \kappa } P^\kappa_{ 000010}
+\frac{ 32 (1 + 2 \kappa) (1 + 12 \kappa) }{ (1 + 7 \kappa) (1 + 11 \kappa) (2 + 11 \kappa) }P^\kappa_{ 100000}\,,\\
&&P^\kappa_{010000}\times P^\kappa_{010000}=
  P^\kappa_{ 020000}
+\frac{ 2 }{ 1 + \kappa } P^\kappa_{ 000100}
+\frac{ 8 (1 + 2 \kappa) }{ (1 + 3 \kappa) (1 + 5 \kappa) } P^\kappa_{ 100001}
+\frac{ 12 (5 + 84 \kappa + 255 \kappa^2 + 160 \kappa^3) }{ (1 + 5 \kappa)^2 (1 + 11 \kappa) (3 + 11 \kappa)  } P^\kappa_{ 010000}\\
&&\quad+\frac{ 144 (1 + 2 \kappa) (1 + 3 \kappa) (1 + 5 \kappa) (1 + 12 \kappa) }{ 
  (1 + 7 \kappa) (1 + 8 \kappa) (1 + 11 \kappa)^2 (2 + 11 \kappa) } P^\kappa_{ 000000}\,,\\
&&P^\kappa_{100000}\times P^\kappa_{001000}=
  P^\kappa_{ 101000}
+\frac{ 3 }{ 1 + 2 \kappa } P^\kappa_{ 000100}
+\frac{ 16 (1 + 2 \kappa) (1 + 8 \kappa) }{ (1 + 5 \kappa) (1 + 7 \kappa) (2 + 7 \kappa) } P^\kappa_{ 
  100001}
+\frac{ 15 (1 + \kappa) (1 + 3 \kappa) (1 + 11 \kappa) }{ (1 + 4 \kappa) (1 + 5 \kappa)^2 (1 + 7 \kappa) } P^\kappa_{ 
  010000}\,,\\
&&P^\kappa_{010000}\times P^\kappa_{001000}=
  P^\kappa_{ 011000}
+\frac{ 4 }{ 1 + 3 \kappa } P^\kappa_{ 100010}
+\frac{ 20 (1 + \kappa) (1 + 8 \kappa) }{ (1 + 4 \kappa) (1 + 7 \kappa) (2 + 7 \kappa) } P^\kappa_{ 
  010001}
+\frac{ 16 (1 + 2 \kappa) }{ (1 + 5 \kappa) (1 + 7 \kappa) } P^\kappa_{ 200000}\\
&&\quad+\frac{ 20 (1 + 2 \kappa) (1 + 9 \kappa) (3 + 46 \kappa+ 71 \kappa^2 - 8 \kappa^3) }{ 
  (1 + \kappa) (1 + 4 \kappa)^2 (1 + 7 \kappa) (1 + 11 \kappa) (3 + 11 \kappa) } P^\kappa_{ 001000}
+\frac{ 80 (1 + \kappa) (1 + 2 \kappa) (1 + 3 \kappa) (1 + 12 \kappa) }{ 
  (1 + 4 \kappa) (1 + 5 \kappa) (1 + 7 \kappa)^2 (2 + 11 \kappa) } P^\kappa_{ 000001}\,,\\
&&P^\kappa_{001000}\times P^\kappa_{001000}=
  P^\kappa_{ 002000}
+\frac{ 2 }{ 1 + \kappa } P^\kappa_{ 100100}
+\frac{ 6 (1 + \kappa) }{ (1 + 2 \kappa) (1 + 3 \kappa) } P^\kappa_{ 010010}
+\frac{ 8 (1 + 2 \kappa) }{ (1 + 3 \kappa) (1 + 5 \kappa) } P^\kappa_{ 200001}\\
&&\quad+\frac{ 4 (9 + 113 \kappa + 305 \kappa^2 + 231 \kappa^3 - 18 \kappa^4) }{ 
  (1 + \kappa) (1 + 3 \kappa)^2 (1 + 7 \kappa) (3 + 7 \kappa) } P^\kappa_{ 001001}
+\frac{ 120 (1 + \kappa) (1 + 2 \kappa) (2 + 33 \kappa + 56 \kappa^2) }{ 
  (1 + 4 \kappa) (2 + 5 \kappa) (1 + 7 \kappa) (2 + 7 \kappa) (3 + 10 \kappa) } P^\kappa_{ 110000}\\
&&\quad+\frac{ 80 (1 + \kappa) (1 + 2 \kappa) (1 + 3 \kappa) (1 + 8 \kappa) }{ 
  (1 + 4 \kappa) (1 + 5 \kappa) (1 + 7 \kappa)^2 (2 + 7 \kappa) } P^\kappa_{ 000002}
+\frac{ 80 (1 + 8 \kappa) (3 + 56 \kappa+176 \kappa^2 + 108 \kappa^3 -63 \kappa^4) }{ 
  (1 + 4 \kappa)^2 (1 + 7 \kappa)^2 (2 + 7 \kappa) (3 + 11 \kappa) } P^\kappa_{ 000010}\\
&&\quad+\frac{ 160 (1 + \kappa)^2 (1 + 2 \kappa) (1 + 3 \kappa) (1 + 9 \kappa) (1 + 12 \kappa) }{ 
  (1 + 4 \kappa)^2 (1 + 5 \kappa) (1 + 7 \kappa)^3 (2 + 11 \kappa) } P^\kappa_{ 100000}\,,\\
&&P^\kappa_{100000}\times P^\kappa_{000100}=
  P^\kappa_{ 100100}
+\frac{ 4 }{ 1 + 3 \kappa } P^\kappa_{ 010010}
+\frac{ 6 (1 + \kappa) (1 + 7 \kappa) }{ (1 + 3 \kappa)^2 (1 + 5 \kappa) } P^\kappa_{ 001001}\\
&&\quad+\frac{ 30 (1 + \kappa) (1 + 2 \kappa) (1 + 8 \kappa) (2 + 11 \kappa) }{ 
  (1 + 4 \kappa) (1 + 5 \kappa)^2 (2 + 7 \kappa) (3 + 10 \kappa) } P^\kappa_{ 110000}
+\frac{ 30 (1 + \kappa) (1 + 2 \kappa) (2 + 5 \kappa) (1 + 8 \kappa) (1 + 9 \kappa) }{ 
  (1 + 4 \kappa)^2 (1 + 5 \kappa)^2 (2 + 7 \kappa) (3 + 11 \kappa) } P^\kappa_{ 000010}\,,\\
&&P^\kappa_{010000}\times P^\kappa_{000100}=
  P^\kappa_{ 010100}
+\frac{ 3 }{ 1 + 2 \kappa } P^\kappa_{ 001010}
+\frac{ 12 (1 + \kappa) (1 + 6 \kappa) }{ (1 + 3 \kappa) (1 + 5 \kappa) (2 + 5 \kappa) } P^\kappa_{ 
  110001}+\frac{ 30 (1 + \kappa) (1 + 2 \kappa) (2 + 11 \kappa) }{ (1 + 4 \kappa) (1 + 5 \kappa)^2 (3 + 10 \kappa) } P^\kappa_{ 
  020000}\\
  &&\quad
+\frac{ 20 (1 + \kappa) (1 + 2 \kappa) (1 + 8 \kappa) }{ (1 + 3 \kappa) (1 + 4 \kappa) (1 + 5 \kappa) 
   (2 + 7 \kappa) } P^\kappa_{ 101000}
+\frac{ 20 (1 + \kappa) (1 + 2 \kappa) (1 + 8 \kappa) }{ (1 + 3 \kappa) (1 + 4 \kappa) (1 + 5 \kappa) 
   (2 + 7 \kappa) } P^\kappa_{ 000011}\\
   &&\quad
+\frac{ 72 (1 + 7 \kappa) (1 + 22 \kappa +115 \kappa^2 + 87 \kappa^3 - 45 \kappa^4) }{ 
  (1 + 3 \kappa)^2 (1 + 5 \kappa)^2 (1 + 11 \kappa) (4 + 11 \kappa) } P^\kappa_{ 000100}
+\frac{ 144 (1 + \kappa)^2 (1 + 2 \kappa) (2 + 5 \kappa) (1 + 8 \kappa)^2 }{ 
  (1 + 3 \kappa) (1 + 5 \kappa)^3 (2 + 7 \kappa)^2 (3 + 11 \kappa) } P^\kappa_{ 100001}\\
  &&\quad
+\frac{ 60 (1 + \kappa) (1 + 2 \kappa)^2 (1 + 3 \kappa) (2 + 5 \kappa) (1 + 8 \kappa) (1 + 9 \kappa) (1 + 11 \kappa) }{  (1 + 4 \kappa)^2 (1 + 5 \kappa)^4 (1 + 7 \kappa) (2 + 7 \kappa) (3 + 11 \kappa) } P^\kappa_{ 010000}\,,\\
&&P^\kappa_{001000}\times P^\kappa_{000100}=
  P^\kappa_{ 001100}
+\frac{ 3 }{ 1 + 2 \kappa }  P^\kappa_{ 110010}
+\frac{ 10 (1 + \kappa) }{ (1 + 3 \kappa) (1 + 4 \kappa) } P^\kappa_{ 020001}
+\frac{ 12 (1 + \kappa) (1 + 6 \kappa) }{ (1 + 3 \kappa) (1 + 5 \kappa) (2 + 5 \kappa) } P^\kappa_{ 
  101001}
\\&&\quad+\frac{ 10 (1 + \kappa) }{ (1 + 3 \kappa) (1 + 4 \kappa) } P^\kappa_{ 000020}
+\frac{ 72 (1 + 6 \kappa) (1 +11 \kappa + 13 \kappa^2 - 5 \kappa^3) }{ 
  (1 + 2 \kappa) (1 + 5 \kappa) (2 + 5 \kappa) (1 + 7 \kappa) (3 + 7 \kappa) } P^\kappa_{ 000101}
\\&&\quad+\frac{ 20 (1 + \kappa) (1 + 2 \kappa) (1 + 8 \kappa) }{ (1 + 3 \kappa) (1 + 4 \kappa) (1 + 5 \kappa) 
   (2 + 7 \kappa) } P^\kappa_{ 210000}
+\frac{ 48 (1 + \kappa)^2 (1 + 2 \kappa) (1 + 8 \kappa) }{ (1 + 3 \kappa)^2 (1 + 5 \kappa)^2 (2 + 7 \kappa)  } P^\kappa_{ 100002}
\\&&\quad+\frac{ 6 (36 + 1134 \kappa + 12624 \kappa^2 + 65771 \kappa^3 + 172189 \kappa^4 + 224179 \kappa^5 + 
     127295 \kappa^6 + 17700 \kappa^7) }{ 
  (1 + 2 \kappa) (1 + 3 \kappa) (1 + 5 \kappa)^2 (2 + 5 \kappa)^2 (1 + 7 \kappa) (3 + 8 \kappa) } P^\kappa_{ 
  011000}
\\&&\quad+\frac{ 24 (1 + \kappa) (42 + 1024 \kappa + 7069 \kappa^2 + 17092 \kappa^3 + 13653 \kappa^4 + 720 \kappa^5) }{ 
  (1 + 3 \kappa)^2 (1 + 5 \kappa)^2 (3 + 7 \kappa) (3 + 8 \kappa) (4 + 11 \kappa) } P^\kappa_{ 10001
   0}
\\&&\quad+\frac{ 480 (1 + \kappa)^2 (1 + 2 \kappa)^2 (1 + 6 \kappa) (1 + 8 \kappa) }{ 
  (1 + 4 \kappa) (1 + 5 \kappa)^3 (1 + 7 \kappa) (2 + 7 \kappa) (3 + 11 \kappa) } P^\kappa_{ 200000}
\\&&\quad+\frac{ 60 (1 + \kappa) (1 + 2 \kappa) (1 + 8 \kappa) 
   (32 + 842 \kappa + 6313 \kappa^2 + 16912 \kappa^3 + 16251 \kappa^4 + 3330 \kappa^5) }{ 
  (1 + 3 \kappa) (1 + 4 \kappa) (1 + 5 \kappa)^2 (2 + 5 \kappa) (1 + 7 \kappa) (2 + 7 \kappa) (3 + 10 \kappa) 
   (4 + 11 \kappa) } P^\kappa_{ 010001}
\eeqrn
\beqrn
\\&&\quad+\frac{ 60 (1 + \kappa) (1 + 2 \kappa) (2 + 5 \kappa) (1 + 8 \kappa) (1 + 9 \kappa) 
   (3 +40 \kappa + 39 \kappa^2 - 18 \kappa^3) }{ 
  (1 + 3 \kappa)^2 (1 + 4 \kappa)^2 (1 + 5 \kappa)^2 (1 + 7 \kappa) (2 + 7 \kappa) (3 + 11 \kappa) } P^\kappa_{ 
  001000}
\\&&\quad+\frac{ 480 (1 + \kappa)^2 (1 + 2 \kappa)^2 (1 + 3 \kappa) (2 + 5 \kappa) (1 + 8 \kappa) (1 + 9 \kappa) 
   (1 + 12 \kappa) }{ (1 + 4 \kappa)^2 (1 + 5 \kappa)^3 (1 + 7 \kappa)^2 (2 + 7 \kappa) (2 + 11 \kappa) 
   (3 + 11 \kappa) } P^\kappa_{ 000001}\,,\\
&&P^\kappa_{000100}\times P^\kappa_{000100}=
  P^\kappa_{ 000200}
+\frac{ 2 }{ 1 + \kappa } P^\kappa_{ 011010}
+\frac{ 6 (1 + \kappa) }{ (1 + 2 \kappa) (1 + 3 \kappa) } P^\kappa_{ 002001}
+\frac{ 6 (1 + \kappa) }{ (1 + 2 \kappa) (1 + 3 \kappa) } P^\kappa_{ 100020}\\
&&\quad
+\frac{ 6 (1 + \kappa) }{ (1 + 2 \kappa) (1 + 3 \kappa) } P^\kappa_{ 120001}
+\frac{ 20 (1 + \kappa) (1 + 2 \kappa) }{ (1 + 3 \kappa) (1 + 4 \kappa) (1 + 5 \kappa) } P^\kappa_{ 
  030000}
+\frac{ 24 (1 +7 \kappa - 2 \kappa^2) }{ (1 + 2 \kappa) (1 + 5 \kappa) (3 + 5 \kappa) } P^\kappa_{ 
  100101}
\\&&\quad+\frac{ 36 (1 + \kappa)^2 (1 + 13 \kappa + 16 \kappa^2) }{ 
  (1 + 2 \kappa)^2 (1 + 3 \kappa) (1 + 5 \kappa) (3 + 7 \kappa) } P^\kappa_{ 010011}
+\frac{ 36 (1 + \kappa)^2 (1 + 13 \kappa + 16 \kappa^2) }{ 
  (1 + 2 \kappa)^2 (1 + 3 \kappa) (1 + 5 \kappa) (3 + 7 \kappa) } P^\kappa_{ 111000}
\\&&\quad+\frac{ 36 (1 + \kappa) (40  +1064 \kappa + 7172 \kappa^2 + 16301 \kappa^3 + 13138 \kappa^4 +940 \kappa^5 - 
     1800 \kappa^6) }{ 5 (2 + \kappa) (1 + 2 \kappa)^3 (1 + 5 \kappa)^2 (2 + 5 \kappa) (4 + 9 \kappa) } P^\kappa_{ 
  010100}
\\&&\quad+\frac{ 48 (1 + \kappa)^2 (1 + 2 \kappa) (1 + 6 \kappa) }{ (1 + 3 \kappa)^2 (1 + 5 \kappa)^2 (2 + 5 \kappa)  } P^\kappa_{ 200002}
+\frac{ 144 (1 + \kappa) (1 + 6 \kappa) (1 + 11 \kappa + 13 \kappa^2 - 5 \kappa^3) }{ 
  (1 + 3 \kappa)^2 (1 + 5 \kappa)^2 (2 + 5 \kappa) (3 + 7 \kappa) } P^\kappa_{ 001002}
\\&&\quad+\frac{ 144 (1 + \kappa) (1 + 6 \kappa) (1 + 11 \kappa + 13 \kappa^2 - 5 \kappa^3) }{ 
  (1 + 3 \kappa)^2 (1 + 5 \kappa)^2 (2 + 5 \kappa) (3 + 7 \kappa) } P^\kappa_{ 200010}
\\&&\quad+\frac{ 2 (195 + 5540 \kappa + 49198 \kappa^2 + 163456 \kappa^3 + 239715 \kappa^4 + 157964 \kappa^5 + 
     57452 \kappa^6 + 26320 \kappa^7) }{ (1 + 2 \kappa)^4 (1 + 5 \kappa)^2 (3 + 7 \kappa) (5 + 11 \kappa) } P^\kappa_{  001010}
\\&&\quad+E P^\kappa_{ 000003}
+FP^\kappa_{ 110001}
+GP^\kappa_{ 000011}
+E P^\kappa_{ 300000}
\\&&\quad+\frac{ 180 (1 + \kappa)^2 (1 + 2 \kappa) (90 + 2499 \kappa + 31155 \kappa^2 + 193684 \kappa^3 + 611355 \kappa^4 + 
     972155 \kappa^5 + 708750 \kappa^6 + 171000 \kappa^7) }{ 
  (1 + 3 \kappa) (1 + 4 \kappa) (1 + 5 \kappa)^4 (2 + 5 \kappa)^2 (3 + 8 \kappa) (3 + 10 \kappa) (5 + 11 \kappa) } P^\kappa_{ 
  020000}
\\&&\quad+G P^\kappa_{ 101000}
+ H P^\kappa_{ 000100}
+ I P^\kappa_{ 100001}\\
&&\quad+\frac{ 2160 (1 + \kappa)^2 (1 + 2 \kappa)^2 (2 + 5 \kappa) (1 + 8 \kappa) (1 + 9 \kappa) 
   (1 + 22 \kappa +115 \kappa^2 + 87 \kappa^3 - 45 \kappa^4) }{ 
  (1 + 3 \kappa) (1 + 4 \kappa)^2 (1 + 5 \kappa)^6 (2 + 7 \kappa) (3 + 11 \kappa) (4 + 11 \kappa) } P^\kappa_{ 
  010000}
\\&&\quad+\frac{ 4320 (1 + \kappa)^2 (1 + 2 \kappa)^3 (1 + 3 \kappa)^2 (2 + 5 \kappa) (1 + 9 \kappa) (1 + 12 \kappa) }{ 
  (1 + 4 \kappa)^2 (1 + 5 \kappa)^3 (1 + 7 \kappa)^2 (2 + 7 \kappa) (1 + 11 \kappa) (2 + 11 \kappa) (3 + 11 \kappa)  } P^\kappa_{ 000000}\,,\\
&&P^\kappa_{100000}\times P^\kappa_{000010}=
  P^\kappa_{ 100010}
+\frac{ 5 }{ 1 + 4 \kappa } P^\kappa_{ 010001}
+\frac{ 10 (1 + \kappa) (1 + 9 \kappa) }{ (1 + 4 \kappa)^2 (1 + 7 \kappa) } P^\kappa_{ 001000}
+\frac{ 32 (1 + 2 \kappa) (1 + 3 \kappa) (1 + 12 \kappa) }{ (1 + 5 \kappa) (1 + 7 \kappa)^2 (2 + 11 \kappa)  } P^\kappa_{ 000001}\,,\\
&&P^\kappa_{001000}\times P^\kappa_{000010}=
  P^\kappa_{ 001010}
+\frac{ 4 }{ 1 + 3 \kappa } P^\kappa_{ 110001}
+\frac{ 15 (1 + \kappa) }{ (1 + 4 \kappa) (1 + 5 \kappa) } P^\kappa_{ 020000}
+\frac{ 20 (1 + \kappa) (1 + 8 \kappa) }{ (1 + 4 \kappa) (1 + 7 \kappa) (2 + 7 \kappa) } P^\kappa_{ 
  101000}
\\&&\quad+\frac{ 20 (1 + \kappa) (1 + 8 \kappa) }{ (1 + 4 \kappa) (1 + 7 \kappa) (2 + 7 \kappa) } P^\kappa_{ 
  000011}
+\frac{ 6 (3 + 40 \kappa + 39 \kappa^2 - 18 \kappa^3) }{ (1 + 2 \kappa) (1 + 3 \kappa)^2 (1 + 7 \kappa) } P^\kappa_{ 000100}\\
&&\quad+\frac{ 8 (1 + 2 \kappa) (48 + 1342 \kappa + 11893 \kappa^2 + 41323 \kappa^3 + 59235 \kappa^4 + 31311 \kappa^5) }{  (1 + 3 \kappa) (1 + 5 \kappa) (1 + 7 \kappa)^2 (2 + 7 \kappa)^2 (3 + 11 \kappa) } P^\kappa_{ 100001}
\\&&\quad+\frac{60 (1 + 2 \kappa) (1 + 3 \kappa) (1 + 9 \kappa) (3 +46 \kappa +71 \kappa^2 - 8 \kappa^3) }{ 
  (1 + 4 \kappa)^2 (1 + 5 \kappa)^2 (1 + 7 \kappa)^2 (3 + 11 \kappa) } P^\kappa_{ 010000}
+\frac{ 432 (1 + \kappa) (1 + 2 \kappa) (1 + 3 \kappa)^2 (1 + 12 \kappa) }{ 
  (1 + 5 \kappa) (1 + 7 \kappa)^2 (1 + 8 \kappa) (1 + 11 \kappa) (2 + 11 \kappa) } P^\kappa_{ 000000}\,,\\
&&P^\kappa_{100000}\times P^\kappa_{000001}=
  P^\kappa_{ 100001}
+\frac{ 6 }{ 1 + 5 \kappa } P^\kappa_{ 010000}
+\frac{ 27 (1 + 3 \kappa) }{ (1 + 8 \kappa) (1 + 11 \kappa) } P^\kappa_{ 000000}\,,
\eeqrn
where the coefficients $E, F, G, H$ and $I$ are such that
\beqrn
&&E\,(1 + 3 \kappa) (1 + 4 \kappa) (1 + 5 \kappa)^2 (1 + 7 \kappa) (2 + 7 \kappa)= 160 (1 + \kappa)^2 (1 + 2 \kappa)^2 (1 + 8 \kappa)\,,\\
&&F\,(1 + 2 \kappa) (1 + 3 \kappa) (2 + 3 \kappa) (1 + 5 \kappa)^3 (2 + 5 \kappa)^2 (3 + 5 \kappa) (3 + 7 \kappa)^2 (5 + 11 \kappa) =144 (1 + \kappa)^2(270 + 10965 \kappa \\
&&\qquad+ 166113 \kappa^2 + 1237287 \kappa^3 + 5078136 \kappa^4 + 12177475 \kappa^5 + 17282049 \kappa^6 + 13976605 \kappa^7 + 5700600 \kappa^8 + 818500 \kappa^9) \,,\\
 && G=\frac{ 720 (1 + \kappa)^2 (1 + 2 \kappa) (1 + 8 \kappa) 
   (8 + 208 \kappa + 1312 \kappa^2 + 1877 \kappa^3 + 360 \kappa^4 - 300 \kappa^5) }{ 
  (2 + \kappa) (1 + 3 \kappa) (1 + 4 \kappa) (1 + 5 \kappa)^3 (2 + 5 \kappa) (2 + 7 \kappa) (3 + 8 \kappa) 
   (4 + 11 \kappa) }\,,\\ 
&&H\,\frac{(1 + 3 \kappa)^2 (1 + 5 \kappa)^4 (2 + 5 \kappa)^2 (1 + 7 \kappa) (3 + 7 \kappa) 
   (3 + 8 \kappa) (4 + 11 \kappa)}{24 (1 + \kappa)}=  444 + 21566 \kappa + 436658 \kappa^2 + 4716853 \kappa^3 \\
      &&\quad + 29111132 \kappa^4 + 
     102644506 \kappa^5 + 195972356 \kappa^6 + 176806835 \kappa^7   +45083850 \kappa^8 - 6894000 \kappa^9 + 
     10935000 \kappa^{10}\,, \\
 &&I\,(1 + 3 \kappa)^2 (1 + 5 \kappa)^5 (2 + 7 \kappa)^2 (3 + 8 \kappa) (3 + 11 \kappa) 
   (4 + 11 \kappa) =864 (1 + \kappa)^2 (1 + 2 \kappa) (1 + 8 \kappa)
   (16 + 626 \kappa + 8775 \kappa^2 + 55745 \kappa^3\\
 &&\qquad+ 172984 \kappa^4 + 268299 \kappa^5 
   + 193845 \kappa^6 + 
     48150 \kappa^7) .
\eeqrn

\end{document}